\documentclass[aps,prl,twocolumn,superscriptaddress,amsmath,amssymb,footinbib,longbibliography]{revtex4-1}
\usepackage[english]{babel}
\usepackage{amssymb,amsbsy,amsmath}
\usepackage{graphicx}
\usepackage{dcolumn}
\usepackage{bm}
\usepackage{subfigure}
\usepackage{color}
\usepackage{textcomp}
\usepackage[colorlinks,bookmarks=false,citecolor=blue,linkcolor=red,urlcolor=blue]{hyperref}

\newcommand{\op}[1]{%
    \fontdimen12\textfont3=2pt\fontdimen12\scriptfont3=1.4pt%
    \!\null\mathop{\vphantom{#1}\smash{#1}}\limits_{\sim}\null\!}

\newcommand{\figref}[1]{Fig.~\protect\ref{#1}}

\newcommand{\kagome}{kagom\'e}

\newcommand {\Mrel} {\mathcal{M}/\mathcal{M}_{\text{sat}}}
\begin{document}
\title{Magnon crystallization in the  \kagome\ lattice antiferromagnet}

\author{J\"urgen Schnack}
\email{jschnack@uni-bielefeld.de}
\affiliation{Fakult\"at f\"ur Physik, Universit\"at Bielefeld, Postfach 100131, D-33501 Bielefeld, Germany}
\author{J\"org Schulenburg}
\affiliation{Universit\"atsrechenzentrum, Universit\"at Magdeburg, D-39016 Magdeburg, Germany}
\author{Andreas Honecker}
\email{andreas.honecker@cyu.fr}
\affiliation{Laboratoire de Physique Th\'eorique et Mod\'elisation, CNRS UMR 8089,
CY Cergy Paris Universit\'e, F-95302 Cergy-Pontoise Cedex, France}
\author{Johannes Richter}
\email{Johannes.Richter@physik.uni-magdeburg.de}
\affiliation{Institut f\"ur Physik, Universit\"at Magdeburg, P.O. Box 4120, D-39016 Magdeburg, Germany}
\affiliation{Max-Planck-Institut f\"{u}r Physik Komplexer Systeme,
        N\"{o}thnitzer Stra{\ss}e 38, D-01187 Dresden, Germany}

\date{\today}

\begin{abstract}
We present numerical evidence for the crystallization of magnons
below the saturation field at non-zero temperatures for the highly frustrated spin-half
\kagome\ Heisenberg antiferromagnet.
This phenomenon can be traced back to the existence of independent localized magnons or equivalently
flat-band multi-magnon states. We present a loop-gas description of these localized magnons
and a phase diagram of this transition, thus providing
information for which magnetic fields and temperatures magnon crystallization
can be observed experimentally.
The emergence of a finite-temperature continuous transition to a
magnon-crystal is expected to be generic for spin models in dimension
$D>1$ where flat-band multi-magnon ground states break translational
symmetry.
\end{abstract}


\maketitle

\emph{Introduction.}---Strongly correlated electronic spin
systems may possess unusual and thus attractive properties such
as magnetization curves characterized by sequences
of magnetization plateaus with possible crystallization of
magnons as reported for Cd-kapellasite recently 
\cite{ONO:NC19}. This is of course a consequence of the
intricate nature of their many-body eigenstates
\cite{Bal:N10,Sta:RPP15,MeB:CRP16,SaB:RPP17}, which, however,
for, e.g., Hubbard as well as Heisenberg models under special
circumstances can express itself as destructive interference
that ``can lead to a disorder-free localization of 
particles'' \cite{RKB:PRB18}. For translationally invariant  
systems this automatically yields flat bands in the
single-particle energy spectrum, i.e., in one-magnon space in the
case of spin Hamiltonians
\cite{Mie:JPA91,Tas:PRL92,SSR:EPJB01,SHS:PRL02,BlN:EPJB03,RSH:JPCM04,ZhT:PRB04,ZhT:PTPS05}. 
Today, flat-band physics is investigated in several areas of
physics, and many interesting phenomena that are related to flat
bands have been found, see, e.g.,
Refs.~\cite{HuA:PRB10,PRS:CRP13,BeL:IJMPB13,LFB:PRB13,DRM:IJMP15,LAF:APX18}.
Flat-band systems can also be created using, e.g., cold atoms in
optical lattices \cite{JGT:PRL12,SOW:PRL12} or by employing
photonic lattices \cite{VCM:PRL15,MSC:PRL15,BGJ:PRL16}. 

Among the flat-band systems, the highly frustrated quantum
antiferromagnets (AFMs) play a particular role as possible
solid-state realizations. There is a large variety of one-, two-,
and three-dimensional lattices, where at high magnetic fields
the lowest band of one-magnon excitations 
above the ferromagnetic vacuum is completely flat
\cite{DeR:EPJB06,DRH:LTP07}. 
These flat-band antiferromagnets exhibit several exotic features
near saturation, such as a macroscopic magnetization jump at the
saturation field \cite{SHS:PRL02}, a magnetic-field driven
spin-Peierls instability \cite{RDS:PRL04}, a
finite residual entropy at the saturation 
field \cite{ZhT:PRB04,DeR:PRB04,ZhT:PTPS05},
a very strong magnetocaloric effect
\cite{ZhH:JSM04,ZhT:PTPS05,DeR:EPJB06}, and an additional
low-temperature maximum of the 
specific heat signaling the appearance of an additional
low-energy scale \cite{DeR:EPJB06}.

\begin{figure}[t]
\centering
\includegraphics*[clip,width=0.80\columnwidth]{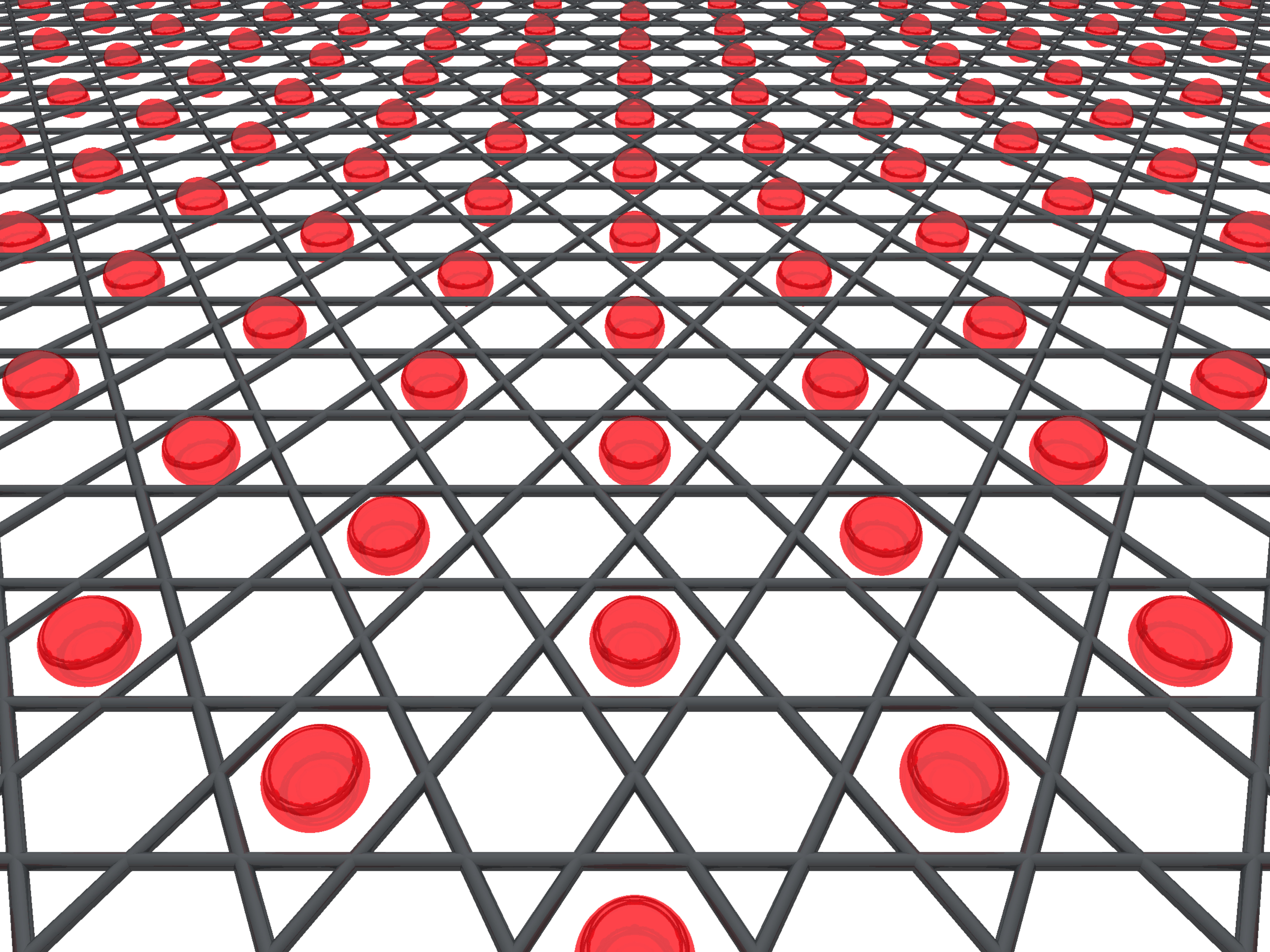}
\caption{Sketch of the crystal of localized magnons (of minimal
size) on the \kagome\ lattice antiferromagnet. These localized
magnons (red discs) are superpositions of spin flips of spins
residing at the vertices of the confining basic hexagons of the
\kagome\ lattice. 
}
\label{magnon_crystal-f-a}
\end{figure}

The focus of the present Letter is on a prominent example of a
flat-band spin system, 
the spin-half \kagome\ Heisenberg antiferromagnet
(KHAF), that is a celebrated paradigm of
highly frustrated quantum  magnetism
\cite{Bal:N10,Sta:RPP15,MeB:CRP16,SaB:RPP17}.
The corresponding Hamiltonian is given by 
\begin{eqnarray}
\label{E-2-1}
\op{H}
&=&
J\;
\sum_{\{i<j\}}\;
\op{\vec{s}}_i \cdot \op{\vec{s}}_j
+ g \mu_B\, B \; \sum_{i}\;
\op{s}^z_i
\ ,
J > 0
\ ,
\end{eqnarray}
where the first term models the Heisenberg exchange between
spins at nearest neighbor sites $i$ and $j$ and the second term
provides the Zeeman splitting in an external magnetic field.

In addition to the widely debated character of the spin-liquid
ground state (GS), the intriguing magnetization process of the KHAF
has attracted much attention
\cite{Hid:JPSJ01,SHS:PRL02,HSR:JP04,RDS:PRL04,ZhT:PRB04,DeR:PRB04,CGH:PRB05,ZhT:PTPS05,DeR:EPJB06,NSH:NC13,CDH:PRB13,NaS:JPSJ18,SSR:PRB18,CRL:SB18,MSS:PRB18,ONO:NC19,ALA:PRB19}.
The magnetization exhibits plateaus at certain fractions of the
saturation magnetization, namely at $\Mrel=3/9=1/3$, $5/9$, $7/9$
and likely also at  $\Mrel=1/9$ \cite{NSH:NC13,CDH:PRB13}.
In contrast to the semiclassical $\Mrel=1/3$ plateau in the  triangular-lattice Heisenberg
antiferromagnet, see, e.g., \cite{ChG:JPCM91,Hon:JPCM99,FZS:JPCM09},
the \kagome\ plateau states are quantum valence-bond states
\cite{RDS:PRL04,ZhT:PRB04,ZhT:PTPS05,CGH:PRB05,NSH:NC13,CDH:PRB13}.
Moreover, around the $\Mrel=7/9$--plateau the flat lowest one-magnon
band \cite{SHS:PRL02} dominates the low-temperature
physics and leads to the exotic properties mentioned above. 
Interestingly, the $\Mrel=7/9$ plateau state just below the jump to saturation is
a magnon crystal that is the magnetic counterpart of the Wigner crystal
of interacting electrons in two dimensions.  
Since the magnon crystal spontaneously breaks translational symmetry,
a finite-temperature phase transition
is possible.
The challenge is to find appropriate theoretical tools to describe such
a transition for the quantum many-body system at hand.

Remarkably, the very existence of a flat band allows a
semi-rigorous analysis of the low-temperature physics, e.g., for
most of the  one-dimensional flat-band quantum spin systems
including the sawtooth chain 
\cite{DeR:PRB04,ZhH:JSM04,ZhT:PTPS05}
and
also for a few two-dimensional systems, such as the frustrated bilayer
\cite{ADP:PRL16,RKB:PRB18,DKR:PRB10}
as well as the Tasaki lattice \cite{MHM:PRL12}.
Such a semi-rigorous analysis builds on the existence of compact
localized many-magnon states, which form either a massively degenerate GS manifold at the
saturation field $B_{\rm sat}$ or a huge set of low-lying
excitations for $B \lesssim B_{\rm sat}$ and  $B \gtrsim B_{\rm sat}$.
For the KHAF, the compact localized many-magnon states live on non-touching hexagons
\cite{SHS:PRL02}, which can be mapped to hard hexagons on a
triangular lattice
\cite{ZhT:PRB04,DeR:PRB04,ZhT:PTPS05,DeR:EPJB06}.
This situation is depicted in \figref{magnon_crystal-f-a}.

On the experimental side the growing number of \kagome\ compounds is promising
with respect to possible solid-state realizations of the \kagome\ flat-band
physics
\cite{Atw:NM02,MBV:PRL07,BNL:PRB07,OTY:PRB11,HHC:N12,KKT:PRB15,IYN:PRL15,Nor:RMP16,YNI:PRB17}.
Very recently the magnetization process in high field was reported 
for Cd-kapellasite \cite{ONO:NC19}. The authors interpret the
observed plateau states ``as crystallizations of emergent
magnons localized on the hexagon of the \kagome\ lattice''.
We will address the relation to our investigations in the discussion below.

Reliable predictions of the field--temperature regions where the
magnon-crystal phase  exists are useful to stimulate specific experiments.   
However, the semi-rigorous analysis of the flat-band properties
of the  KHAF  based on compact localized many-magnon states, i.e., the hard-hexagon approximation
(HHA) is limited 
because of the existence of a macroscopic number of additional
{\it  non-compact} localized many-magnon states \cite{DRH:LTP07}.
A complete description can be given in terms of
a loop gas (LG) that we elaborate in the Supplemental Material \cite{supplementaryMat}.
Moreover, at non-zero temperature also non-localized eigenstates
influence the thermodynamics of the KHAF.

\emph{Numerical method.}---To investigate the KHAF near the saturation field we present
large-scale finite-temperature Lanczos (FTL) 
studies for finite lattices of $N=27$, $\ldots$ 
$72$ sites, where we have
selected only lattices exhibiting the magnon-crystal plateau at
$\Mrel=7/9$, which excludes $N=42$ discussed in \cite{SSR:PRB18}.
FTL is an unbiased numerical approach by which thermodynamic
quantities are very accurately approximated by means of trace
estimators
\cite{JaP:PRB94,ScW:EPJB10,HaS:EPJB14,ScT:PR17,PRE:COR17,SRS:PRR20}. 
Moreover, the  consideration of six different lattices up to $N=72$ allows
to estimate finite-size effects. For used lattices and technical
details see \cite{supplementaryMat}.
The \kagome\ lattices of $N$ sites correspond to
triangular lattices of $N_{\rm trian}=N/3$ sites.
On symmetry grounds, triangular lattices of
$N_{\rm trian}=9,12,21$, {\em i.e.}, $N=27,36,63$ sites seem to be most appropriate
for our investigation \cite{BLP:PRL92,supplementaryMat}.

\begin{figure}[t!]
\centering
\includegraphics*[clip,width=0.90\columnwidth]{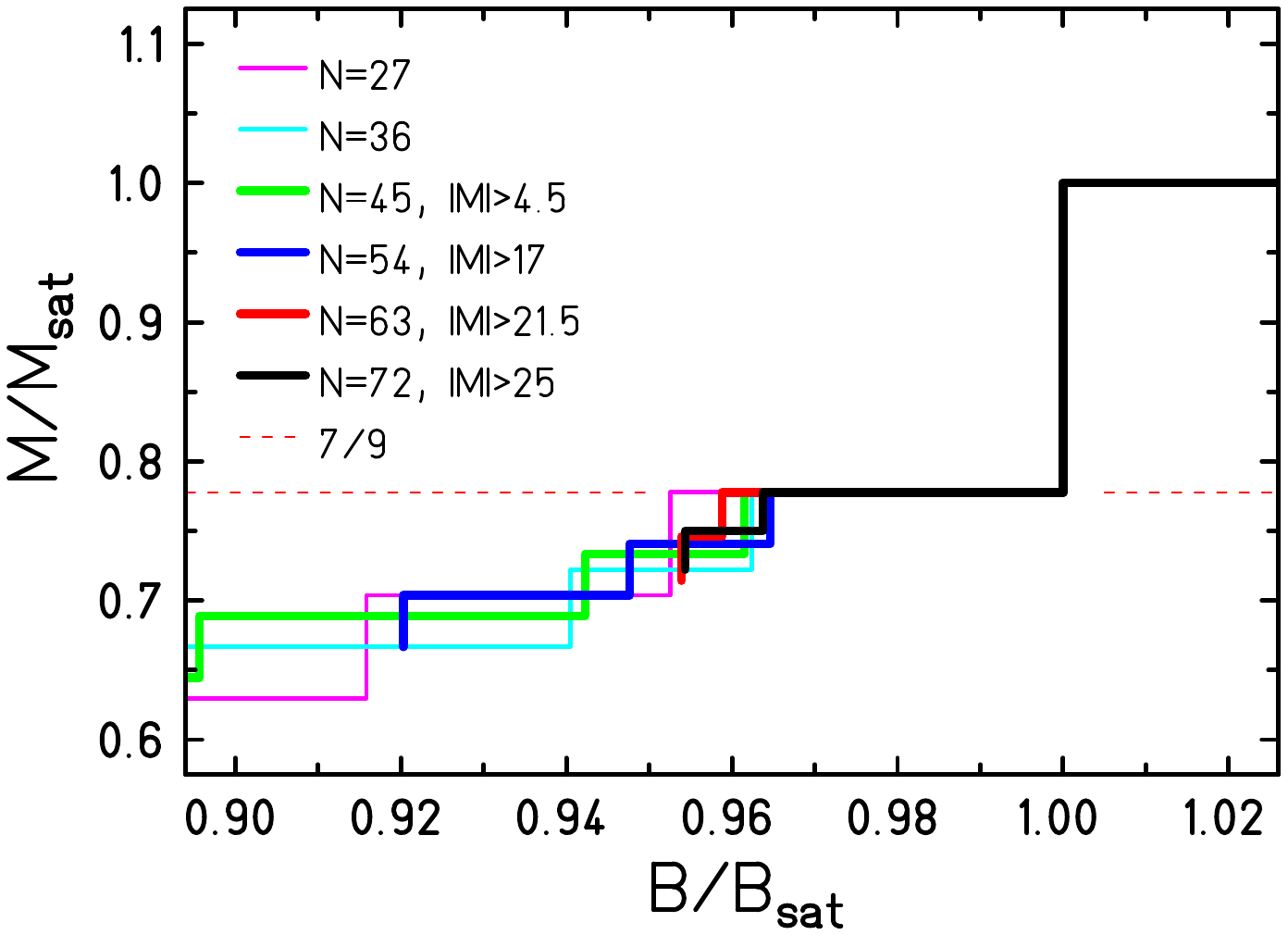}
\caption{Magnetization $\Mrel$: Region of the 7/9 plateau for various
  finite-size realizations of the KHAF.}
\label{plateau-7-9}
\end{figure}

\emph{Results.}---The magnetization
curve around the $7/9$--plateau and the jump to 
saturation are shown in \figref{plateau-7-9}. 
The size-independence of the height of the jump is obvious.
The width of the plateau, i.e., the field region where the magnon-crystal phase can
exist, is about $4\%$ of the saturation field and its
finite-size dependence is weak, cf.\ Ref.~\cite{CDH:PRB13}.
The finite-temperature transition to the magnon-crystal phase can be driven
either by temperature when fixing $B$ in the plateau region or by the
magnetic field when fixing $T$ below the critical temperature $T_c$.
$C(B,T)$ is an appropriate quantity to detect the transition.
For finite lattices the specific heat will not
exhibit a true singularity, rather we may expect a
well-pronounced peak in $C$ that indicates the critical
point. Furthermore, the peak has to become 
sharper with increasing $N$.

\begin{figure}[t!]
\centering
\includegraphics*[clip,width=0.90\columnwidth]{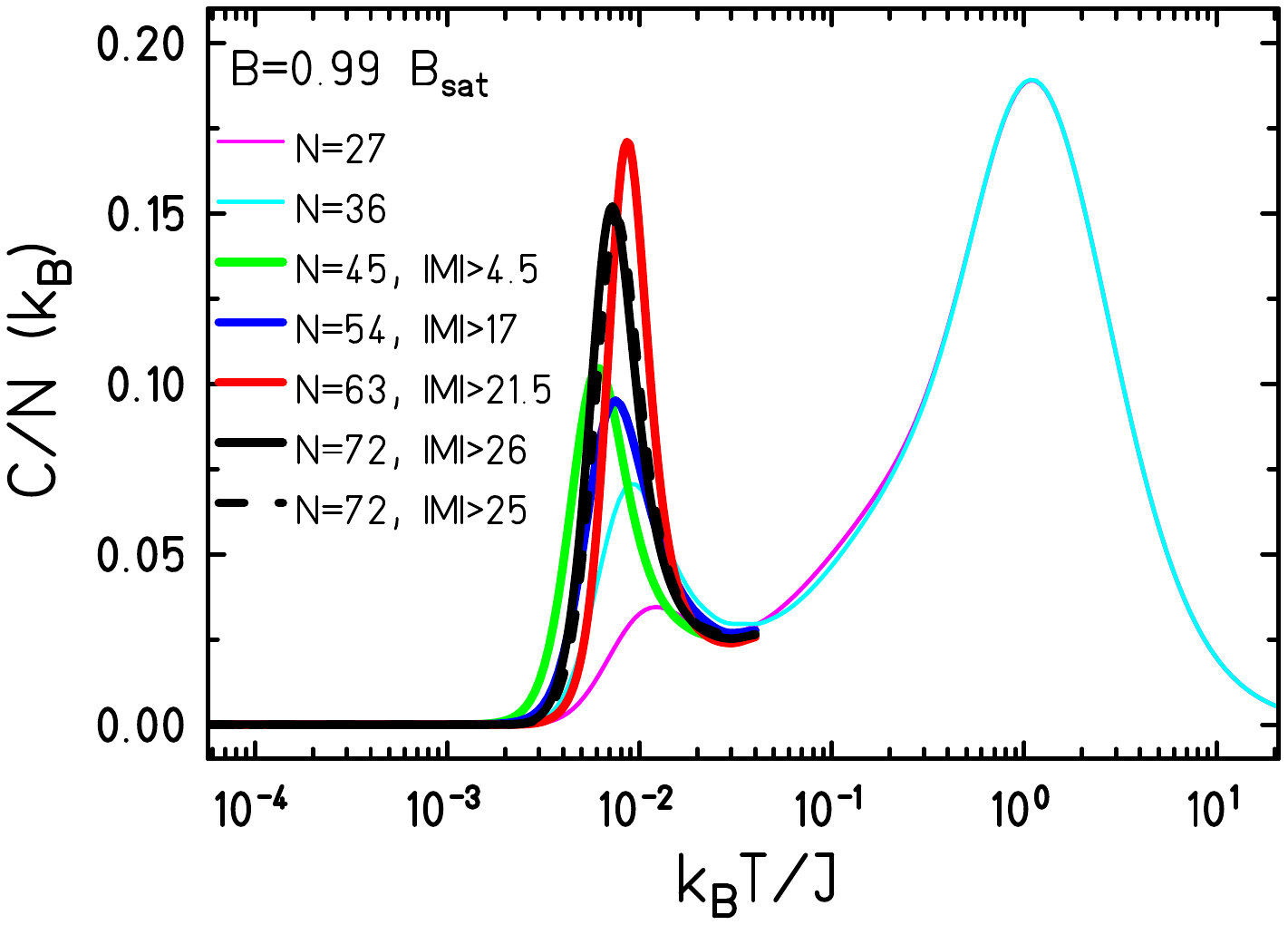}
\caption{
Specific heat for $B=0.99B_{\rm sat}$ for various finite-size realizations of the
KHAF. For $N=45,54,63,72$, where too large Hilbert subspaces had
to be neglected, only the low-temperature part of the specific
heat is displayed; it is virtually correct for all system sizes.}
\label{C-T-099Bsat}
\end{figure}

First, we study the temperature profile $C(T)$ for a magnetic field
slightly below saturation, $B=0.99\,B_{\rm sat}$ (see
\figref{C-T-099Bsat}).
While the influence of $N$ on the peak position $T_{\rm max}$ is rather weak, the
increase of the height $C_{\rm max}$ with growing $N$ is significant and
the peaks are sharpest for $N=63$ and $N=72$.

\begin{figure}[t!]
\centering
\includegraphics*[clip,width=0.90\columnwidth]{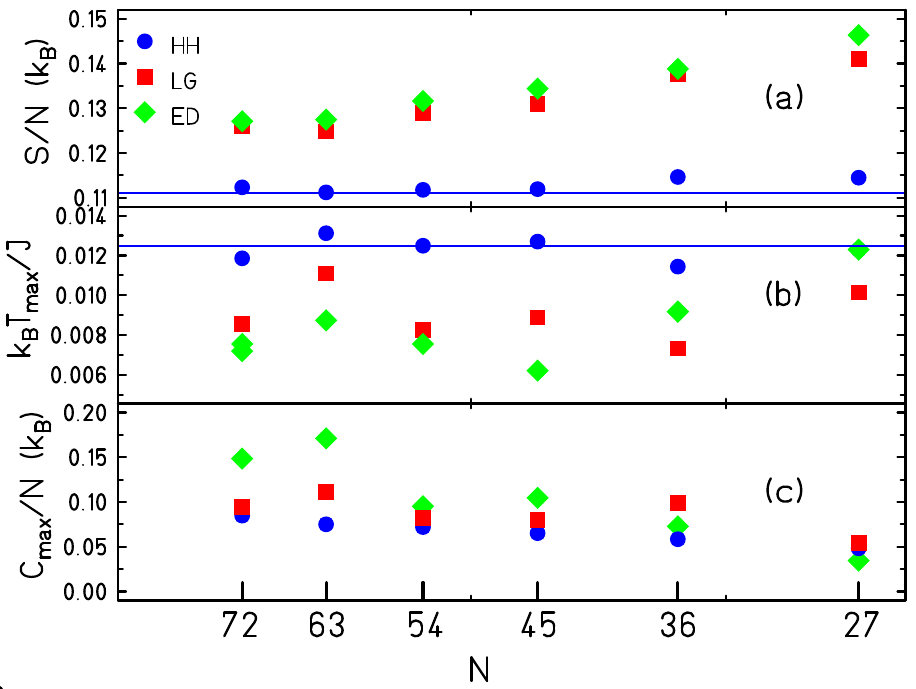}
\caption{System-size dependence of several characteristic quantities
at $B = 0.99\,B_{\rm sat}$
for HH, LG, and ED:
(a) entropy per site $S/N$ associated to the total ground-state degeneracy,
(b) position of the maximum of the specific heat $T_{\max}$, and
(c) value of the maximum of the specific heat per site $C_{\max}/N$.
The horizontal blue lines in panels (a) and (b) show the known thermodynamic
limit for hard hexagons \cite{BaT:JPA80,Bax:JPA80,ZhT:PRB04,ZhT:PTPS05}.
} 
\label{Cmax-N-099Bsat}
\end{figure}

Figure~\ref{Cmax-N-099Bsat} presents a closer look at the results
of Fig.~\ref{C-T-099Bsat} in terms of some characteristic quantities
where we include the HHA \cite{ZhT:PRB04,ZhT:PTPS05} and the LG description
\cite{supplementaryMat}
for comparison. In panel (a) we first present a comparison of the total
ground-state entropy per site. Since hard hexagons are a subset of the
loop configurations that in turn are a subset of the KHA ground states,
the values of $S$ increase correspondingly for a fixed $N$. For the HHA,
the result for the thermodynamic limit is known \cite{BaT:JPA80,ZhT:PRB04,ZhT:PTPS05}
and shown by the horizontal blue line.
We note that the result for $N=63$ within the HHA
approximation is very close to this $N=\infty$ limit. Since the finite-size
effects of the LG and the KHA are very similar to that of the
HHA, we assume that also for these models a system size of
$N\ge 63$ is at least necessary to arrive at trustworthy
results. It is thus a major achievement that by means of FTL
such sizes are accessible.

We observe furthermore that
nested loop configurations do indeed give rise to another macroscopic
contribution to the ground states \cite{DRH:LTP07}, and while this
is approaching the ED result for the KHA, there is yet another
contribution to the ground-state manifold that does not come
from localized magnons and thus cannot be captured by the LG
either. 

Figure~\ref{Cmax-N-099Bsat}(b) displays the size dependence of the position
of the maximum $T_{\max}$ of the specific heat in all three approximations.
The thermodynamic limit of the HHA is again known \cite{Bax:JPA80,ZhT:PRB04,ZhT:PTPS05}
and again shown by the horizontal blue line. The positions for $N \gtrsim 45$ scatter
around this value, and since the finite-size effects of all three approaches
are again similar, we assume the same to be true for the LG and the KHA.
Thus, we conclude that the critical temperature is lowered by the
higher ground-state degeneracy of the LG and the KHA by up to 50\%\ as
compared to the HHA even for a field as close to the saturation field
as $B = 0.99\,B_{\rm sat}$.

Finally, Fig.~\ref{Cmax-N-099Bsat}(c) shows the size dependence
of $C_{\rm max}/N$. 
The range of accessible system sizes and lattice geometries is
too small to reliably extract critical 
exponents, but one does observe a trend of $C_{\max}/N$ to grow with
increasing system size $N$. To be more precise,
the transition is expected to belong to the universality class
of the classical two-dimensional Potts model
\cite{ZhT:PRB04,ZhT:PTPS05} for all three cases. Thus, the
asymptotic behavior of $C_{\rm max}/N$ for large $N$ should be given by
$C_{\rm max}/N \propto N^{(\alpha/2\nu)}$\cite{KiL:PA98} with
critical indices $\alpha=1/3$ and $\nu=5/6$ \cite{Wu:RMP82,NOJ:CMP13}.

\begin{figure}[t!]
\centering
\includegraphics*[clip,width=0.90\columnwidth]{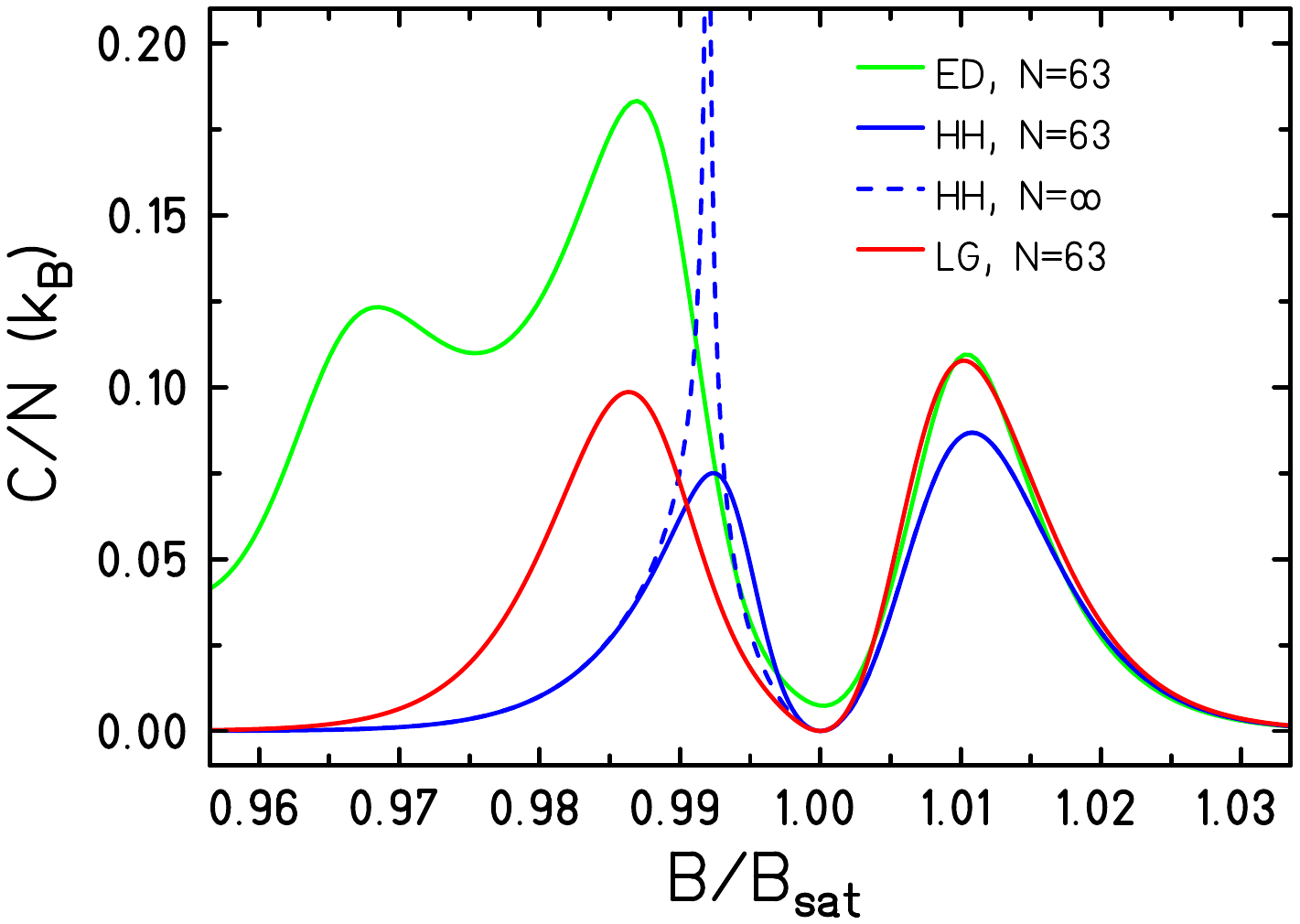}
\caption{Specific heat vs.\ $B$ at $T/J=0.01$ for the KHAF, HH,
and LG with $N=63$ (solid curves) and the thermodynamic limit
of hard hexagons \cite{Bax:JPA80,ZhT:PRB04,ZhT:PTPS05} (dashed curve).}
\label{C-B-fixT-JS}
\end{figure}

Next, we consider the field dependence of the specific heat
for a representative low temperature
$T/J=0.01$, see \figref{C-B-fixT-JS}, where we present data for
$N=63$ (solid curves). For the HHA, we include the result for the
thermodynamic limit $N=\infty$ 
\cite{Bax:JPA80,ZhT:PRB04,ZhT:PTPS05} (dashed curve).
We note that both the HHA and the LG scale with $(B-B_{\rm
  sat})/T$ \cite{ZhT:PRB04,ZhT:PTPS05,DeR:EPJB06}.
There are two peaks left and right of the minimum in $C(B)$ at $B=B_{\rm sat}$ which
are related to the ground states of Fig.~\ref{Cmax-N-099Bsat}(a).
The curves for $N=63$ and $\infty$ of the HHA for the peak of $B>B_{\rm sat}$
are indistinguishable, showing that this is not a phase transition.
The peaks of the LG and and ED for $B>B_{\rm sat}$ are at almost
the same position but higher than for the HHA, and they do not signal a phase
transition either. Remarkably, the LG is very close to the ED result for $B>B_{\rm sat}$,
a fact that can be attributed to the LG reproducing the exact ground-state
degeneracy of the highest sectors of total magnetic quantum
number for the KHAF, see also \cite{supplementaryMat}.

Turning to the region $B<B_{\rm sat}$ of \figref{C-B-fixT-JS},
here the HHA is known to exhibit a phase transition \cite{Bax:JPA80,ZhT:PRB04,ZhT:PTPS05}
whose location is given by the divergence of the $N=\infty$ curve (dashed).
The peaks in the ED and LG curves just below $0.99\,B_{\rm sat}$
should correspond to the same crystallization transition, they
are just rounded off by the finite size and pushed to lower
$B$ compared to the HHA by the larger number of states involved.
In this region, the ED peak is higher than that of the LG.
This difference is not only due to the KHAF having ground states
that have no LG description \cite{supplementaryMat}, but
also due to low-lying excitations. The latter give rise to a second
peak at $0.97\,B_{\rm sat}$ that is present only in the ED data.

\begin{figure}[t!]
\centering
\includegraphics*[clip,width=0.90\columnwidth]{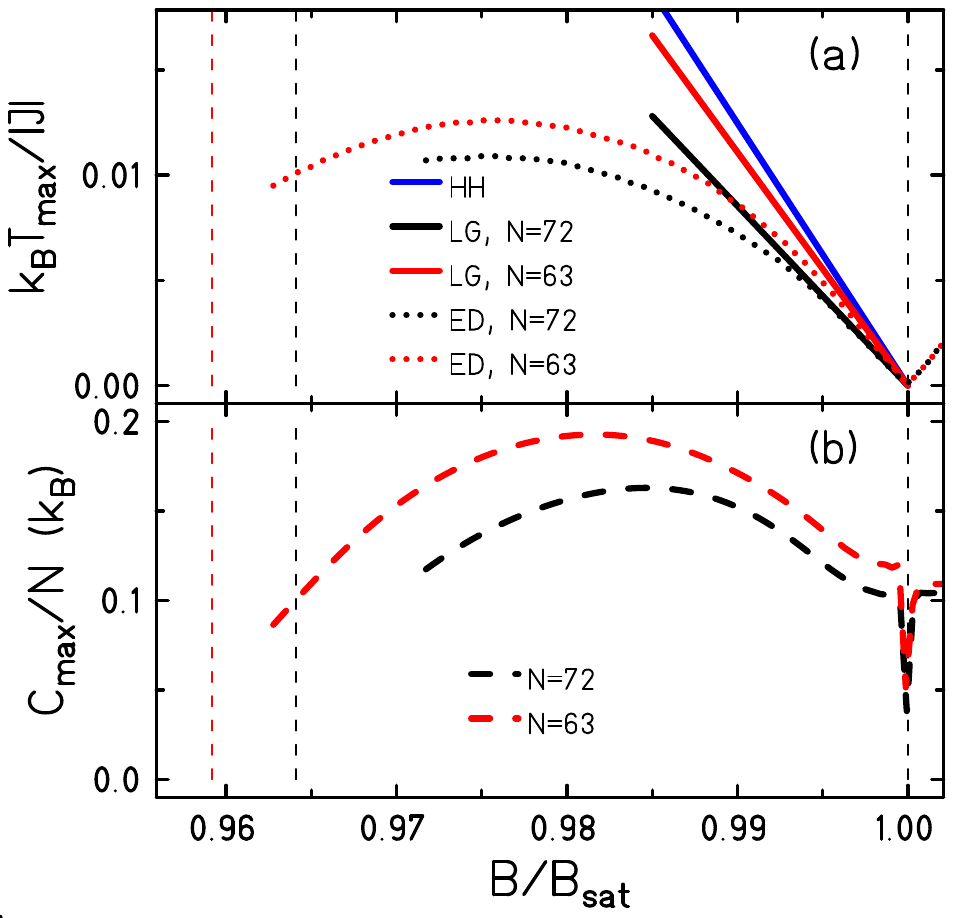}
\caption{Phase diagram: 
  (a) Position $T_{\rm max}$ and (b) height $C_{\rm max}$ of the
  low-$T$ maximum (cf. \figref{C-T-099Bsat}) in dependence on $B$ for $N=63$ and $N=72$
  for fields where the maximum can be
  unambiguously detected. The vertical dashed lines mark the
  repective edges of the magnetization plateau.
}
\label{Cmax}
\end{figure}

To derive a tentative  phase diagram,  we show
in \figref{Cmax}(a) the
position $T_{\rm max}$ of the low-$T$ peak of $C(T)$ vs
$B$ for $N=63$ and $N=72$. We also show the
HHA result $T_c=0.928(1-B/B_{\rm sat})$ for $N=\infty$
\cite{ZhT:PRB04,ZhT:PTPS05} 
and the LG curves for $N=63$ and $72$ (straight lines).
The LG curves are very close to tangential to the corresponding
ED results just below the saturation field, while the HHA yields
a higher transition temperature, as already noticed in the context
of Fig.~\ref{Cmax-N-099Bsat}(b). As $B$ decreases, the ED curves
bend down, and
when approaching the lower endpoint $B_{\rm end}$ of the plateau
(depicted by the vertical lines in \figref{Cmax}) $T_{\rm max}$
decreases and we may expect that it vanishes near $B_{\rm end}$, where
the magnon-crystal ground state disappears.
For finite systems, as approaching
$B_{\rm end}$ the relevant peak in $C(T)$ 
merges with low-$T$ finite-size peaks appearing just below $B_{\rm end}$,
this way masking the true behavior expected for $N\to \infty$.
 
We mention that the general shape of the transition curve
in \figref{Cmax}(a) resembles the phase
diagram of the magnon crystallization of the fully frustrated bilayer AFM
\cite{ADP:PRL16,RKB:PRB18,DKR:PRB10}.
Therefore, we may argue that the shape of this curve is
generic for two-dimensional spin models possessing flat-band
multi-magnon ground states.

The height of the maximum $C_{\rm max}$ of $C(T)$
(supposed to become a power-law singularity for $N\to\infty$) is    
shown in \figref{Cmax}(b) vs $B$ for $N=63$ and $N=72$. 
The shape of these curves is dome-like with a maximum near the
midpoint of the plateau. The unusual behavior at $B=B_{\rm sat}$
is discussed in Ref.~\cite{DeR:EPJB06}.

\emph{Discussion.}---Our FTL data confirm the very existence of 
a low-temperature magnon-crystal phase just below the saturation
field as conjectured by the 
HHA \cite{ZhT:PRB04,ZhT:PTPS05}. However, the $B$--$T$ region
where this phase exists is not properly described by the HHA. Instead
we elaborated a LG description that complements our FTL
investigations. It is very accurate for
$B>B_{\rm sat}$ and still yields a good description just below
$B_{\rm sat}$. Our investigations therefore provide guidance in
which range of field and temperature a magnon-crystal phase is
to be expected.

Coming back to the ``magnon crystallization'' reported in the
experimental 
paper \cite{ONO:NC19}: Here the authors interpret the
observed plateau states ``as crystallizations of emergent
magnons localized on the hexagon of the \kagome\ lattice''.
This concept coincides with the present study for the
$7/9$--plateau, but may differ
for plateaus at smaller magnetization, e.g., at  1/3 and  5/9. 
Although these lower plateaus can be understood 
as magnon crystals formed at $T=0$, it still has to be
investigated whether the physical behavior
for $T>0$ differs from the scenario discussed in this
Letter, since the huge set of flat-band multi-magnon states determining the
low-$T$ thermodynamics near $B_{\rm sat}$ is missing for these
plateaus.

As already discussed by the authors of \cite{ONO:NC19} a real
compound always differs from the idealized theoretical case for
instance due to long-range dipolar or Dzyaloshinskii-Moriya
interactions. In the case of Cd-kapellasite these seem to
stabilize a phase at 10/12 of the saturation
magnetization. However, the structure of this phase
appears to be rather similar to that at 7/9, it therefore served
as a strong motivation to investigate the possibility of a
magnon crystallization phase transition on very general
grounds (and with an idealized Hamiltonian).
The effect of certain anisotropic Hamiltonians on magnon
crystal phases confined to \kagome\ stripes 
is extensively discussed in e.g. \cite{ALA:PRB19}.


\emph{Acknowledgment.}---This work was supported by the Deutsche
Forschungsgemeinschaft (DFG SCHN 615/23-1). Computing time at
the Leibniz Center in Garching is gratefully acknowledged.
The authors are indebted to O.~Derzhko, J.~Stre\v{c}ka, and M.~E.\ Zhitomirsky for
valuable discussions.


%

\end{document}


\title{Supplemental Material for \\
``Magnon crystallization in the  \kagome\ lattice antiferromagnet''}

\author{J\"urgen Schnack}
\affiliation{Fakult\"at f\"ur Physik, Universit\"at Bielefeld, Postfach 100131, D-33501 Bielefeld, Germany}
\author{J\"org Schulenburg}
\affiliation{Universit\"atsrechenzentrum, Universit\"at Magdeburg, D-39016 Magdeburg, Germany}
\author{Andreas Honecker}
\affiliation{Laboratoire de Physique Th\'eorique et Mod\'elisation, CNRS UMR 8089,
CY Cergy Paris Universit\'e, F-95302 Cergy-Pontoise Cedex, France}
\author{Johannes Richter}
\affiliation{Institut f\"ur Physik, Universit\"at Magdeburg, P.O. Box 4120, D-39016 Magdeburg, Germany}
\affiliation{Max-Planck-Institut f\"{u}r Physik Komplexer Systeme,
        N\"{o}thnitzer Stra{\ss}e 38, D-01187 Dresden, Germany}

\date{\today}

\maketitle

\section{Lattices}

\label{sec:latt}

\begin{figure*}[t!]
\centering
\hbox{\raise 2 mm\hbox{\includegraphics[width=0.45\textwidth]{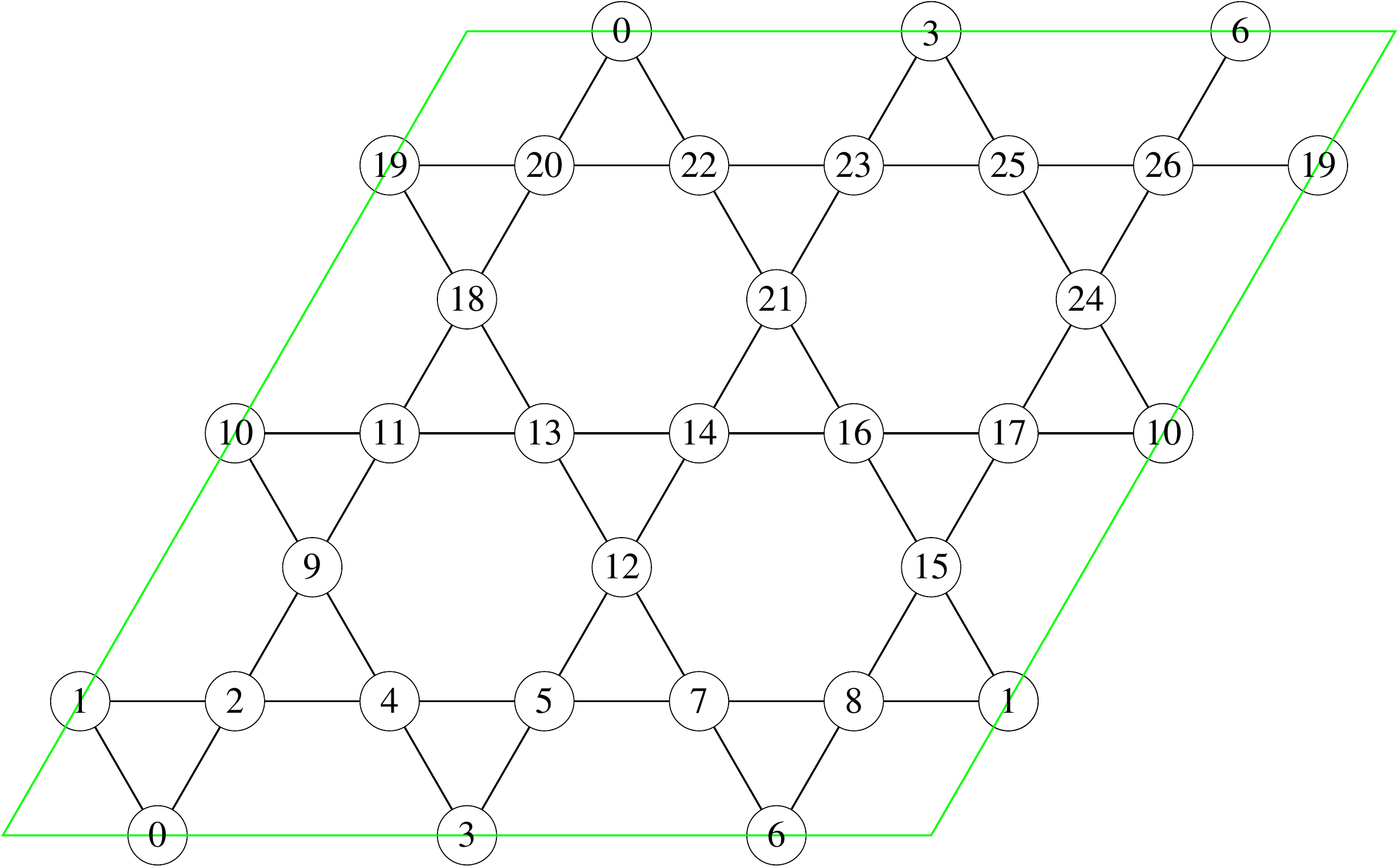}} \qquad
\hbox{\includegraphics[width=0.53\textwidth]{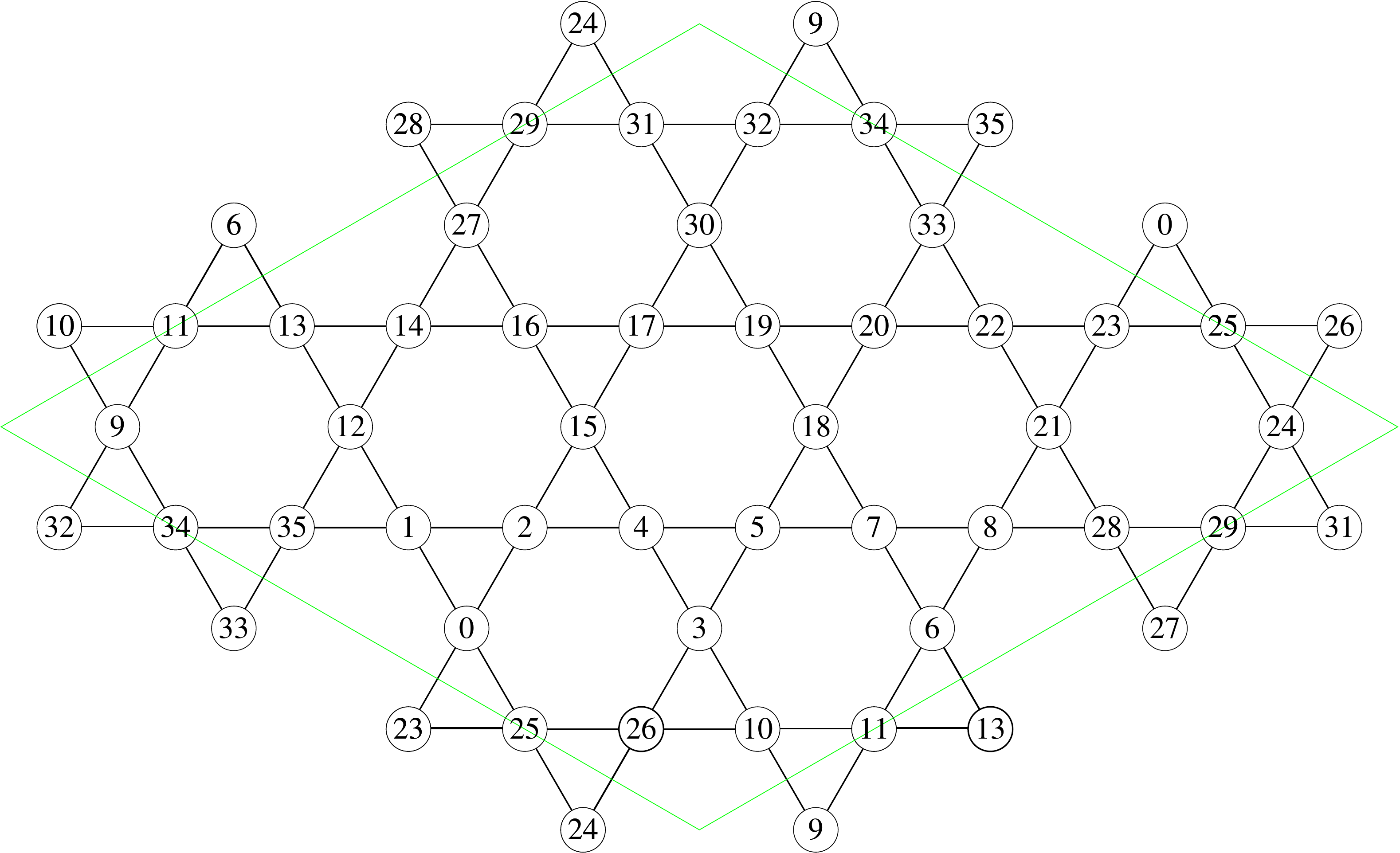}}}
\vspace*{3mm}
\hbox{\raise 8 mm\hbox{\includegraphics[width=0.49\textwidth]{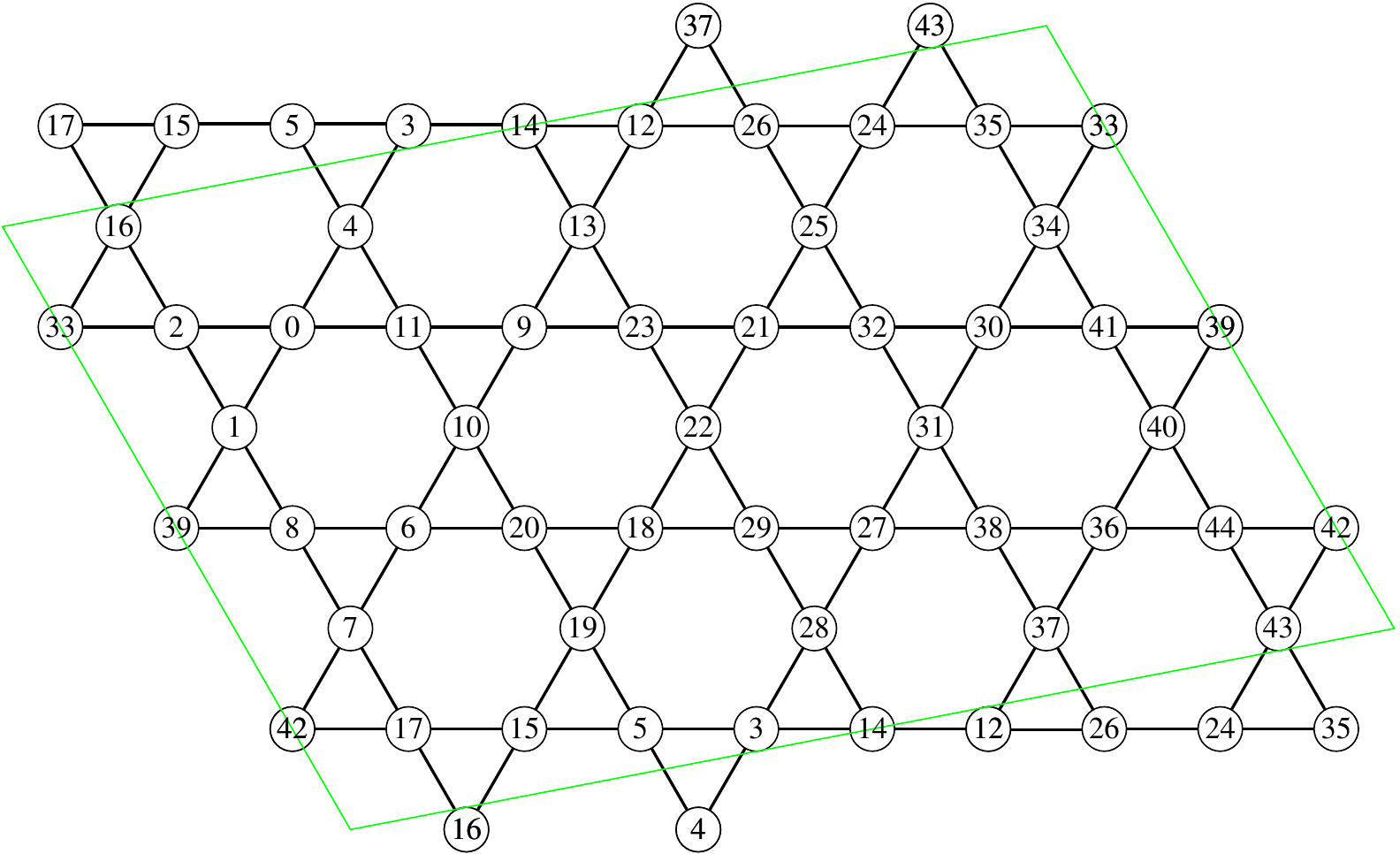}} \qquad
\hbox{\includegraphics[width=0.49\textwidth]{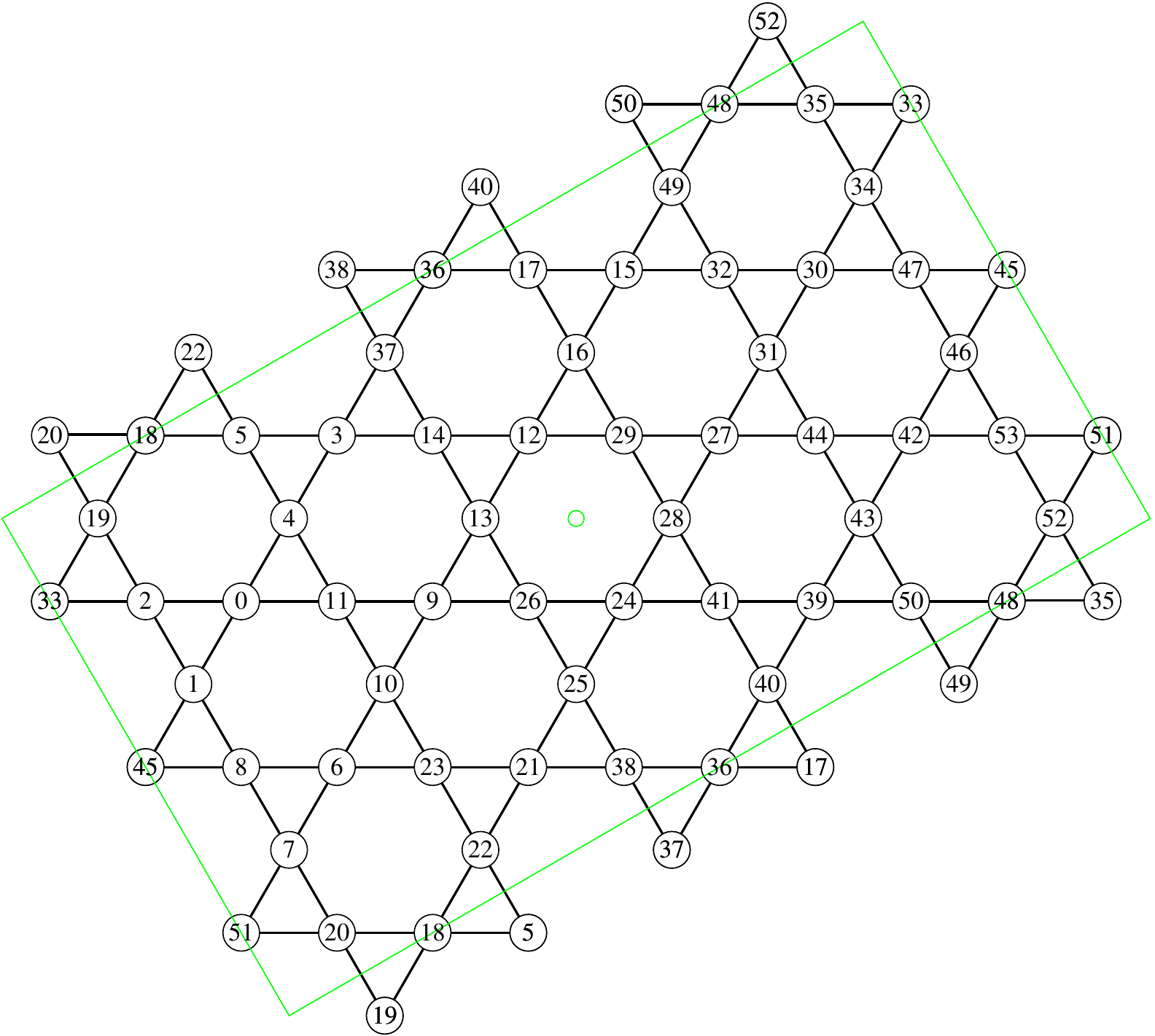}}}
\vspace*{-6mm}
\includegraphics[width=0.41\textwidth]{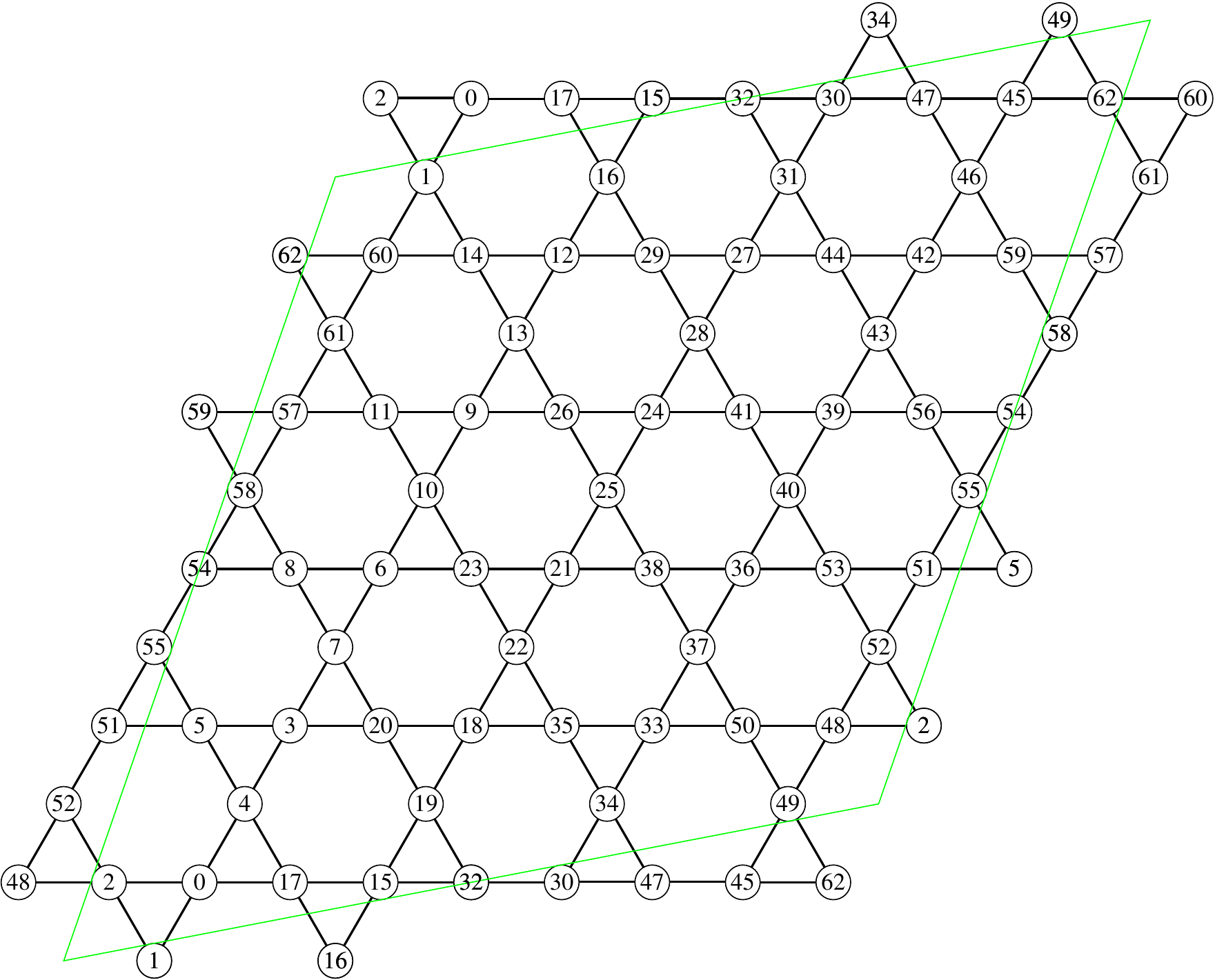}
\hspace*{-0.12\textwidth}
\includegraphics[width=0.69\textwidth]{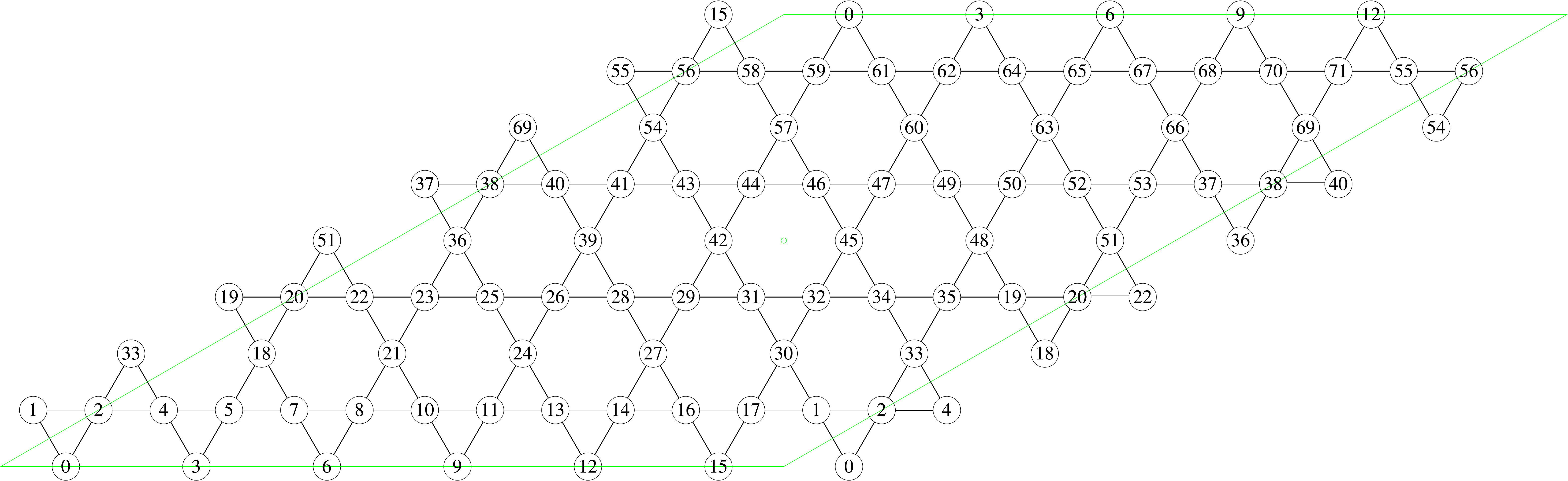}
\caption{Finite \kagome\ lattices with $N=27$, $36$, $45$, $54$, $63$, and $N=72$ sites
from top left to bottom right.}
\label{magnon_crystal-lattice}
\end{figure*}

Figure \ref{magnon_crystal-lattice} shows the lattices employed in the present work.
The green lines indicate periodic boundary conditions.
Firstly, we note that the lattices with $N=27$, $36$, and $63$ share the six-fold
rotational symmetry of the infinite lattice around the center of a hexagon.
Indeed, the \kagome\ lattice can be considered as a triangular lattice of $N_{\rm trian}=N/3$ sites,
decorated by a basis of triangles, and the corresponding effective triangular lattices 
with $N_{\rm trian} = 9$, $12$, and $21$ are known to be favorable from a symmetry point
of view \cite{BLP:PRL92}.


Secondly, we note that the $N=27$, $45$, and $54$ lattices have loops of length 6 wrapping
around the boundaries.

\section{Magnetic quantum numbers}

The eigenstates of the model are characterized by
the magnetic quantum number $M$ belonging to the $z$-component
$\op{s}^z$ of the total spin
and the $\vec{k}$-vector of the translational symmetry.  
While for $N=27$ and $N=36$ we can take into account all sectors of $|M|$,
for $N > 36$ we are restricted to sectors of larger $|M|$: 
 $|M|>9/2$ for $N=45$, $|M|>17$ for $N=54$, $|M|>43/2$ for $N=63$, and
$|M|>26$ for $N=72$, respectively. This restriction is not severe,
since close to the saturation field the eigenstates with small
$|M|$ become excited states with higher energy. Nevertheless,
for $N>36$ we are restricted to low enough
temperatures to avoid substantial contributions of states with
small $|M|$ to the partition function.


\section{ED size scaling}

\begin{figure}[tb!]
\centering
\includegraphics[width=0.99\columnwidth]{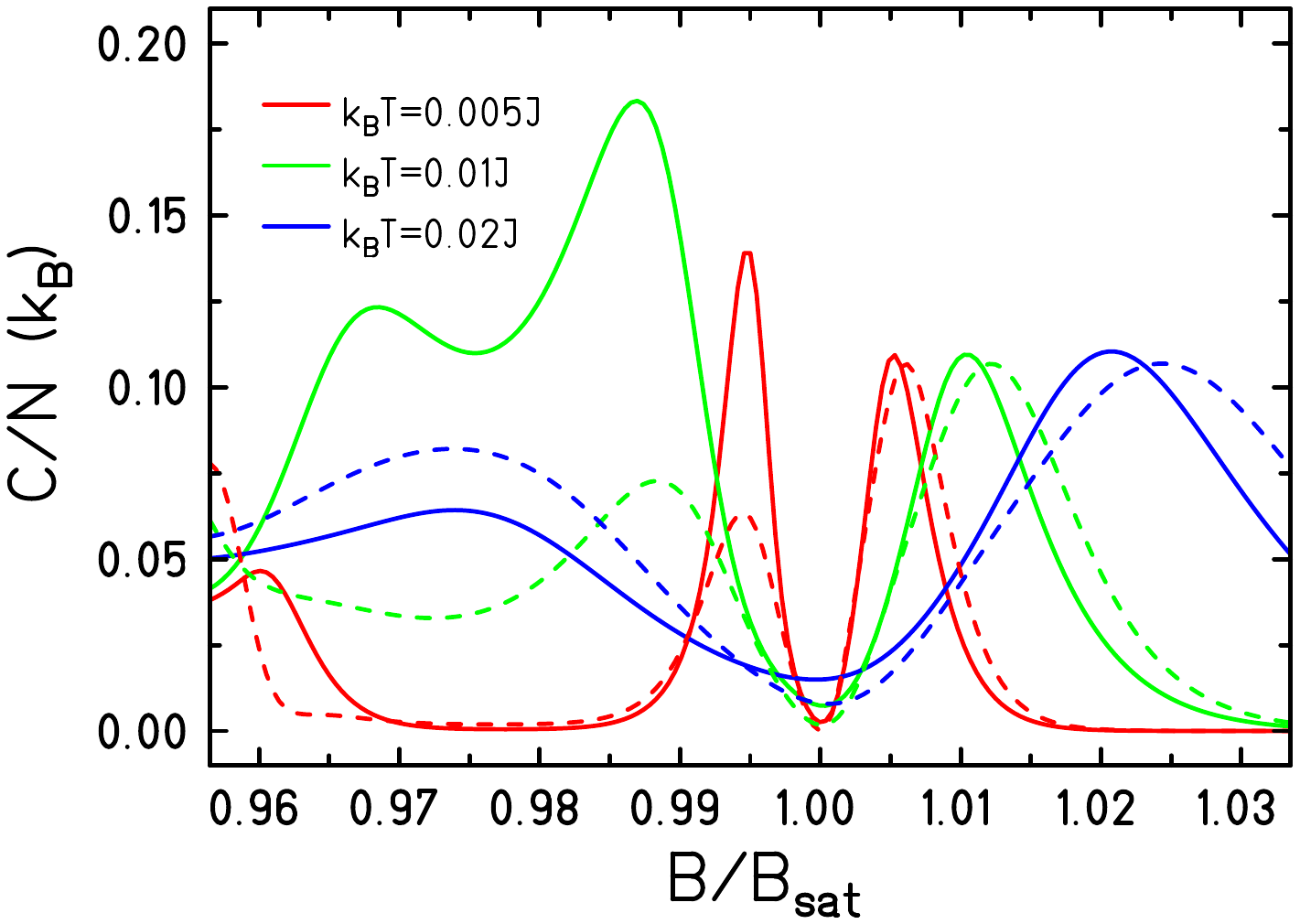}
\caption{Specific heat vs. $B$ at various low temperatures for the KHAF
with $N=63$ (solid curves) and $N=36$ (dashed curves, same color
for same temperature).}
\label{fig:CBold5}
\end{figure}

Here would like to provide an additional figure, complementing Fig.~5 of 
the main text, that shows the field dependence of the specific heat at 
low temperatures $T/J=0.005,0.01,0.02$, see \figref{fig:CBold5}, where we 
present ED data for $N=36$ (dashed) and $N=63$ (solid curves). There are 
two peaks left and right of the minimum in $C(B)$ at $B=B_{\rm sat}$ 
which are related to the huge set of low-lying excitations. The peaks are 
sharp at very low $T/J=0.005$ and become broader with increasing $T$. The 
height of the maximum above $B_{\rm sat}$ is almost identical for $N=36$ 
and $N=63$; it does not correspond to a phase transition 
\cite{DeR:EPJB06,DRH:LTP07}, see also the main text.

\section{Loop-gas description}

One can map the localized magnons states in the high-field
regime of the \kagome\ lattice to a {\em geometric} problem of loop configurations
\cite{SHS:PRL02,ZhT:PRB04,RSH:JPCM04,ZhT:PTPS05,DRH:LTP07}.

\begin{figure}[tb!]
\centering
\includegraphics[width=0.99\columnwidth]{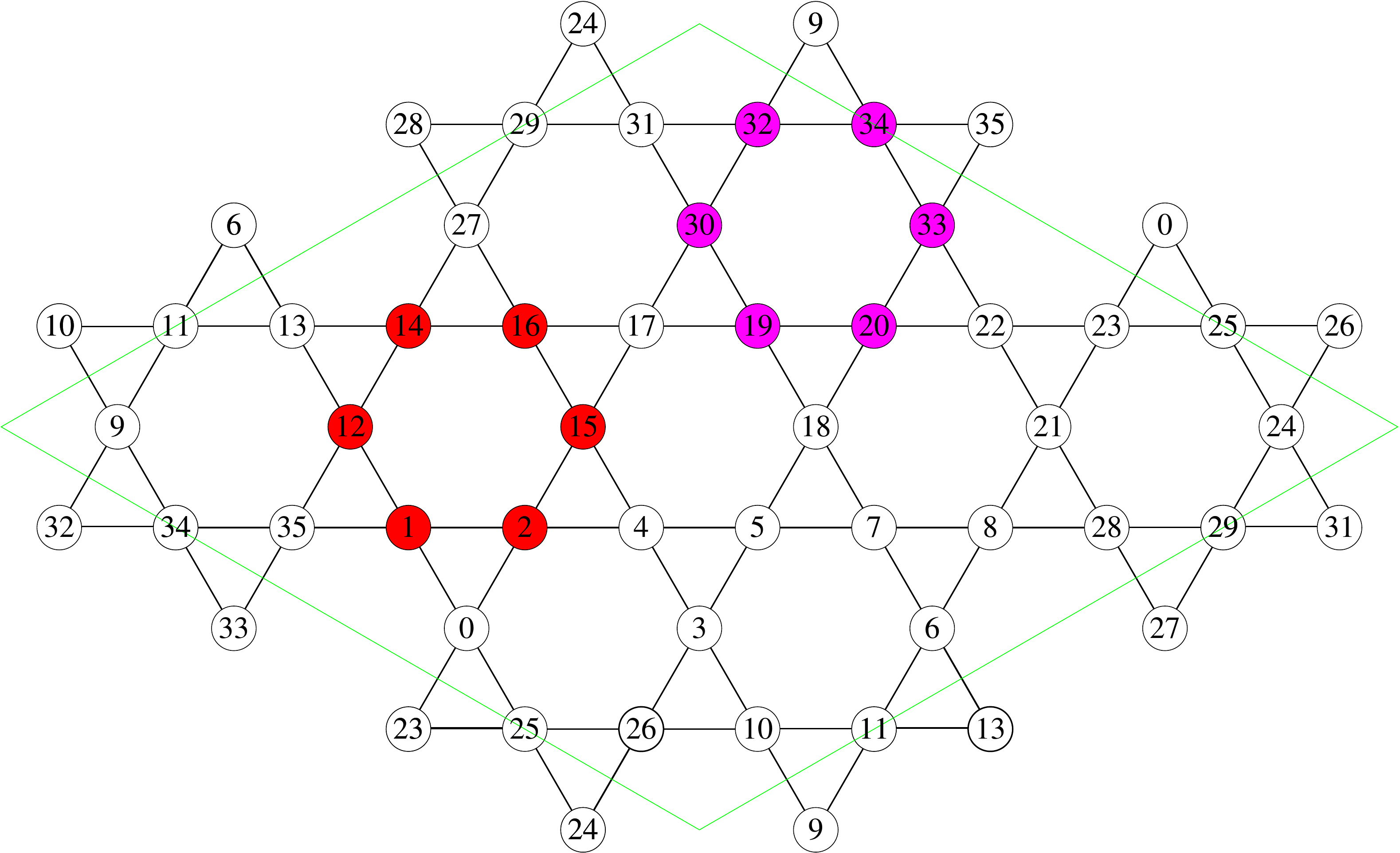} \\[1mm]
\includegraphics[width=0.99\columnwidth]{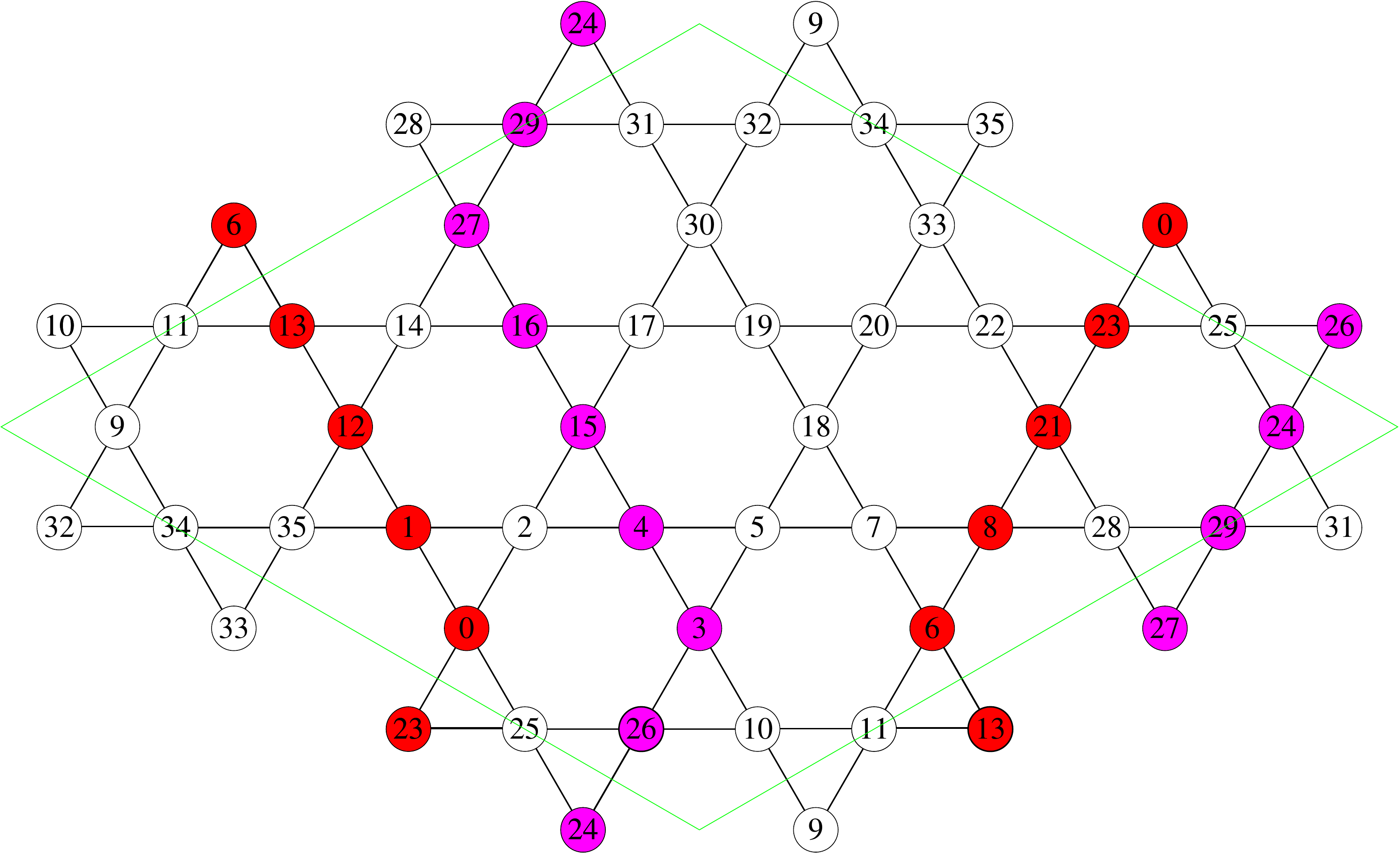} \\[1mm]
\includegraphics[width=0.99\columnwidth]{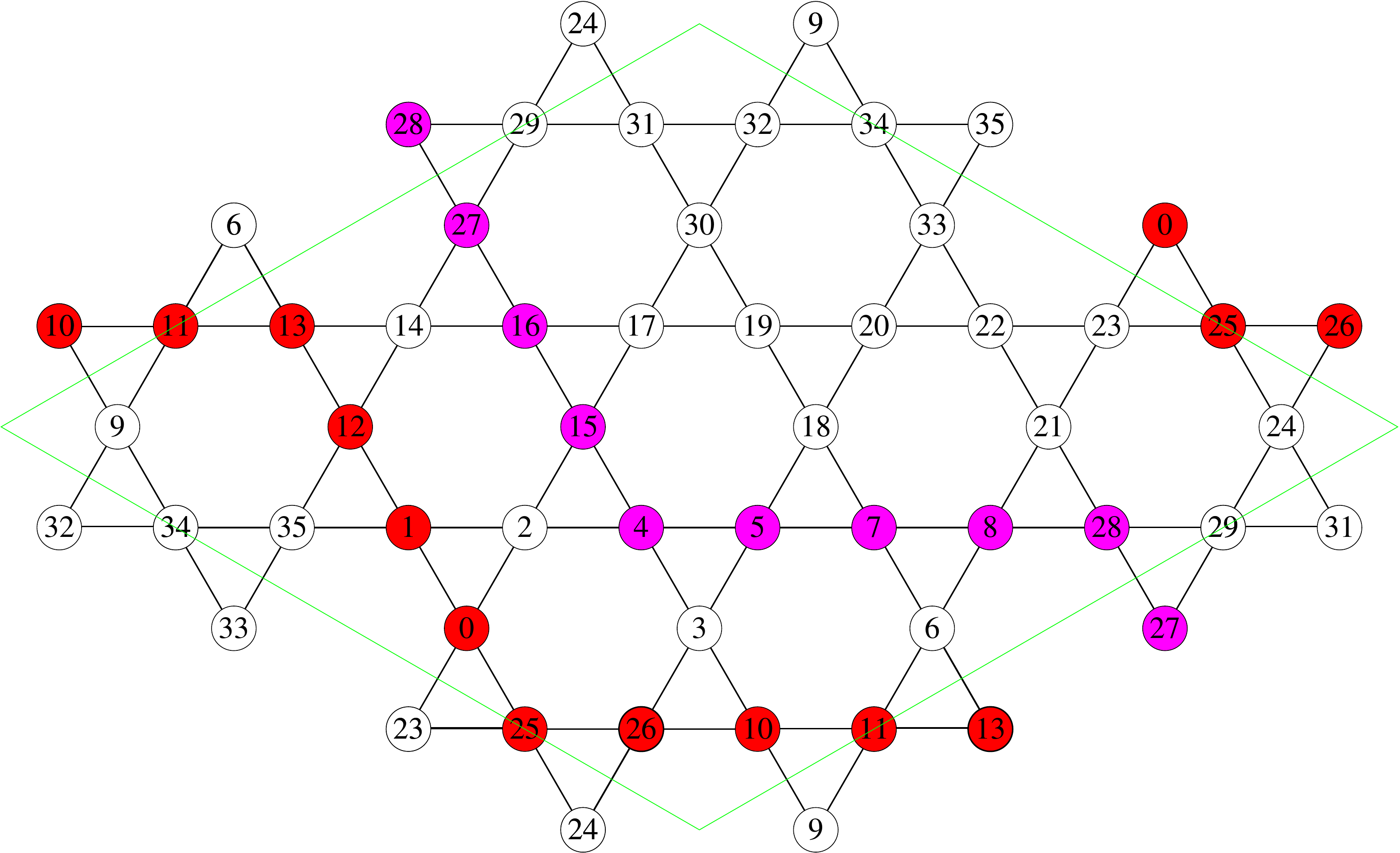}
\caption{Top: Configuration with two minimal localized magnons
(minimal loops) on the $N=36$ \kagome\ lattice.
Center and bottom: Two configurations with two loops winding around the boundary.
The sites belonging to the two loops are shown by the filled red and magenta circles,
respectively.
\label{fig:N36w2}
}
\end{figure}

Let $\state{\uparrow \ldots \uparrow}$ be the ferromagnetically polarized
state of the spin-1/2
Heisenberg antiferromagnet on a \kagome\ lattice with $N$ sites.
Then one can construct exact eigenstates in the sector $S^z=N/2-n_{\ell}$ using
$n_{\ell}$ closed loops and
\begin{equation}
\state{\{\ell_i\}} = \prod_{i=1}^{n_\ell} \left(
\sum_{x_i\in \ell_i} (-1)^{x_i} \, \op{s}_{x_i}^-
\right) \, \ferrostate \, .
\label{eq:loopState}
\end{equation}
Here $(-1)^{x_i}$ stands symbolically for an alternating sign along the loop $\ell_i$.
For illustration, Fig.~\ref{fig:N36w2} shows three configurations
consisting of two loops on the $N=36$ lattice. The top panel
shows a configuration consisting of two loops of minimal length,
{\it i.e.}, compact localized magnons,
corresponding to a configuration with two hard hexagons.

\subsection{Properties of loop configurations}

In order for the states Eq.~(\ref{eq:loopState}) to be exact eigenstates, the
loop configurations must satisfy the following conditions:
\begin{enumerate}
\item The length $\abs{\ell_i}$ of the loop $\ell_i$ must be even in order to accommodate the
   alternating sign of the wave function along it.
\item A ``loop'' of length $2$ turning back on its tail is not allowed since it would leave a single
   flipped spin in two triangles and thus not lead to an exact eigenstate.
\item In each triangle, at most two sites are allowed to be occupied. This is required in order to
   ensure destructive interference when a spin flip wants to propagate outside a loop. This condition
   implies not only that a loop $\ell_i$ cannot approach itself too closely when turning back on
   itself, but also that two loops $\ell_i$ and $\ell_j$ must be separated at least by one free site.
\end{enumerate}
Actually, in order to ensure the exact eigenstate property of Eq.~(\ref{eq:loopState}), exactly
zero or two sites in each triangle must be occupied by a loop.
Indeed, this is already ensured by the above rules.
We further note that, at least for a sufficiently big lattice, the
shortest loops on the \kagome\ lattice have length $\abs{\ell_i}=6$. These include the loops winding around a hexagon,
{\it i.e.},
the compact localized magnon states or
``hard-hexagon'' states. In addition, as mentioned in Sec.~\ref{sec:latt}, one
may also have loops of length 6 winding around a periodic boundary.

The above rules can be implemented on a computer, thus allowing computer enumeration of the loop
configurations for a finite lattice with $N$ sites.

\subsection{Linear relations}

The wave functions Eq.~(\ref{eq:loopState}) are not all linearly independent. For example, if
the configurations
$\{\ell_i\}$ and $\{\ell_i'\}$ differ by a loop $\ell_i'-\ell_i$ only in their
$i$th factor  and $(\ell_1,\ldots,\ell_{i-1},\ell_i'-\ell_i,\ell_{i+1},\ldots,\ell_{n_\ell})$
is also valid loop configuration, then $\state{\{\ell_i\}}$, $\state{\{\ell_i'\}}$, and
$\state{\ell_1,\ldots,\ell_{i-1},\ell_i'-\ell_i,\ell_{i+1},\ldots,\ell_{n_\ell}}$ are
linearly dependent such that we may eliminate, e.g., $\state{\{\ell_i'\}}$ from the spanning
set.

For convenience, we recall what is known about the degeneracies of the ground-state manifold on an
$N$-site \kagome\ lattice \cite{DRH:LTP07}:
\begin{itemize}
\item
The degeneracy is known to be $N/3+1$ in the $n_\ell=1$ sector. This is evident, e.g., from a band picture.
An equivalent counting is $N/3$ hard hexagons, subject to one linear relation plus two winding states.
\item
In the sector corresponding to $n_\ell=2$, the degeneracy is $(N-3)\,(N-6)/18$.
Ref.~\cite{DRH:LTP07} gave the following interpretation of this number: firstly, there are
$N/6\,(N/3-7)$ configurations with two hard hexagons. Secondly, there are $2\,N/3$ independent configurations composed of one hard hexagon and a winding state. Thirdly, there is one further linearly independent state composed of two winding states.
\end{itemize}
The latter statement indicates that a purely geometric picture will be insufficient.
To illustrate this point, we show in the center and bottom panel of Fig.~\ref{fig:N36w2} two double-winding configurations
on an $N=36$ \kagome\ lattice. Since the winding number is conserved by local moves, it is not
possible to deform these two configurations into each other by local moves, or to reduce them
to hard hexagons, neither to one hard hexagon and one winding configuration. Therefore, it will in general be
necessary to investigate the linear relations between the vectors Eq.~(\ref{eq:loopState}).

The hard-hexagon configurations can still be considered to be linearly independent. On the torus created by the periodic boundary conditions, this is not strictly true, but the resulting linear relation can be compensated by adding a winding state.

In general, we determine the linear relations as follows. First, we note that the scalar product of two vectors of the type  Eq.~(\ref{eq:loopState}) can be computed
within the loop representation. It is therefore not necessary to work in the possibly high-dimensional
vector space of spin configurations with $S^z = N/2-n_{\ell}$, but one can in fact perform
a Gram-Schmidt orthogonalization in the abstract space of the states Eq.~(\ref{eq:loopState})
and in this
manner determine the dimension of the vector space spanned by them.

\subsection{Results of computer enumeration}

\begin{table*}[tb!]
\begin{tabular}{ccc|ccc|c}
\multicolumn{3}{c|}{Heisenberg model} &
\multicolumn{3}{c|}{loop gas} &
hard hexagons \\ \hline
$S^z$ & degeneracy & gap $\Delta/J$ & $n_\ell$ & \#\ confs.\
 & \#\ lin.\  indep.
 & \# confs.\ \\
 \hline
$25/2$ & $10$ & $0.63397$ & $1$ & $783$ & $10$ & $9$ \\
$23/2$ & $28$ & $0.20165$ & $2$ & $234$ & $28$ & $9$ \\
$21/2$ & $13$ & $0.04379$ & $3$ & $6$   & $6$  & $3$ \\
\end{tabular}
\caption{Data for the $N=27$ lattice.
For the Heisenberg model, we quote the gap to the lowest excited
state above the degenerate ground-state manifold in the corresponding
sector of $S^z$.
\label{tab:N27}
}
\end{table*}

\begin{table*}[tb!]
\begin{tabular}{ccc|ccc|c}
\multicolumn{3}{c|}{Heisenberg model} &
\multicolumn{3}{c|}{loop gas} &
hard hexagons \\ \hline
$S^z$ & degeneracy & gap $\Delta/J$ & $n_\ell$ & \#\ confs.\
 & \#\ lin.\  indep.
 & \# confs.\ \\
 \hline
$17$ & $13$ & $0.5$     & $1$ & $5442$ & $13$ & $12$ \\
$16$ & $55$ & $0.18159$ & $2$ & $2616$ & $55$ & $30$ \\
$15$ & $71$ & $0.05548$ & $3$ & $130$  & $70$ & $16$ \\
$14$ & $8$  & $0.03363$ & $4$ & $3$    & $3$  & $3$ \\
\end{tabular}
\caption{Data for the $N=36$ lattice. For the $S=1/2$ Heisenberg model
and the number of hard-hexagon configurations compare Ref.~\cite{DRH:LTP07}.
\label{tab:N36}
}
\end{table*}

\begin{table*}[tb!]
\begin{tabular}{ccc|ccc|c}
\multicolumn{3}{c|}{Heisenberg model} &
\multicolumn{3}{c|}{loop gas} &
hard hexagons \\ \hline
$S^z$ & degeneracy & gap $\Delta/J$ & $n_\ell$ & \#\ confs.\
 & \#\ lin.\  indep.
 & \# confs.\ \\
 \hline
$43/2$ & $16$  & $0.25139$ & $1$ & $37221$ & $16$  & $15$ \\
$41/2$ & $91$  & $0.12257$ & $2$ & $23530$ & $91$  & $60$ \\
$39/2$ & $201$ & $0.03478$ & $3$ & $4520$  & $190$ & $60$ \\
$37/2$ & $110$ & $0.01055$ & $4$ & $260$   & $60$  & $15$ \\
$35/2$ & $4$   & $0.01176$ & $5$ & $4$     & $4$   & $3$ \\
\end{tabular}
\caption{Data for the $N=45$ lattice. For the $S=1/2$ Heisenberg model
and the number of hard-hexagon configurations compare Ref.~\cite{DRH:LTP07}.
\label{tab:N45}
}
\end{table*}

\begin{table*}[tb!]
\begin{tabular}{ccc|ccc|c}
\multicolumn{3}{c|}{Heisenberg model} &
\multicolumn{3}{c|}{loop gas} &
hard hexagons \\ \hline
$S^z$ & degeneracy & gap $\Delta/J$ & $n_\ell$ & \#\ confs.\
 & \#\ lin.\  indep.
 & \# confs.\ \\
 \hline
$26$ & $19$  & $0.17712$ & $1$ & $255696$ & $19$  & $18$ \\
$25$ & $136$ & $0.09119$ & $2$ & $195975$ & $136$ & $99$ \\
$24$ & $430$ & $0.02458$ & $3$ & $63036$  & $413$ & $180$ \\
$23$ & $513$ & $0.00901$ & $4$ & $9192$   & $396$ & $99$ \\
$22$ & $119$ & $0.00295$ & $5$ & $384$    & $90$  & $18$ \\
$21$ & $4$   & $0.01237$ & $6$ & $4$      & $4$   & $3$ \\
\end{tabular}
\caption{Data for the $N=54$ lattice. For the $S=1/2$ Heisenberg model
and the number of hard-hexagon configurations compare Ref.~\cite{DRH:LTP07}.
\label{tab:N54}
}
\end{table*}

\begin{table*}[tb!]
\begin{tabular}{ccc|ccc|c}
\multicolumn{3}{c|}{Heisenberg model} &
\multicolumn{3}{c|}{loop gas} &
hard hexagons \\ \hline
$S^z$ & degeneracy & gap $\Delta/J$ & $n_\ell$ & \#\ confs.\
 & \#\ lin.\  indep.
 & \# confs.\ \\
 \hline
$61/2$ & $22$    & $0.29675$ & $1$ & $1927644$ & $22$   & $21$  \\
$59/2$ & $190$   & $0.12841$ & $2$ & $1743441$ & $190$  & $147$ \\
$57/2$ & $785$   & $0.05004$ & $3$ & $481509$  & $784$  & $406$ \\
$55/2$ & $1436$  & $0.02395$ & $4$ & $40656$   & $1288$ & $399$ \\
$53/2$ & $617$   & $0.00629$ & $5$ & $1029$    & $294$  & $105$ \\
$51/2$ & $21$    & $0.01132$ & $6$ & $21$      & $21$   & $21$  \\
$49/2$ & $3$     & $0.04807$ & $7$ & $3$       & $3$    & $3$   \\
\end{tabular}
\caption{Data for the $N=63$ lattice. For the
$S=1/2$ Heisenberg model in the sectors with $S^z \ge 57/2$
and the number of hard-hexagon configurations compare Ref.~\cite{DRH:LTP07},
for the degeneracy of the Heisenberg model in the $S^z=49/2$ sector,
compare Ref.~\cite{CDH:PRB13}.
\label{tab:N63}
}
\end{table*}

\begin{table*}[tb!]
\begin{tabular}{ccc|ccc|c}
\multicolumn{3}{c|}{Heisenberg model} &
\multicolumn{3}{c|}{loop gas} &
hard hexagons \\ \hline
$S^z$ & degeneracy & gap $\Delta/J$ & $n_\ell$ & \#\ confs.\
 & \#\ lin.\  indep.
 & \# confs.\ \\
 \hline
$35$ & $25$   & $0.13397$ & $1$ & $12611908$ & $25$   & $24$   \\
$34$ & $253$  & $0.07558$ & $2$ & $13519854$ & $253$  & $204$  \\
$33$ & $1293$ & $0.02521$ & $3$ & $5961328$  & $1293$ & $752$  \\
$32$ & $3303$ & $0.01013$ & $4$ & $1321624$  & $3207$ & $1218$ \\
$31$ & $3512$ & $0.00462$ & $5$ & $133700$   & $2864$ & $816$  \\
$30$ &        & $0.00333$ & $6$ & $5894$     & $954$  & $212$  \\
$29$ &        &           & $7$ & $120$      & $84$   & $24$   \\
$28$ &        &           & $8$ & $3$        & $3$    & $3$    \\
\end{tabular}
\caption{Data for the $N=72$ lattice.
\label{tab:N72}
}
\end{table*}

Tables \ref{tab:N27}--\ref{tab:N72} summarize results obtained by 
computer enumeration and subsequent Gram-Schmidt orthogonalization for 
the lattices shown in Fig.~\ref{magnon_crystal-lattice}. These tables 
include results for the Heisenberg model and hard hexagons, where for the 
convenience of the reader we also reproduce some data from 
Refs.~\cite{DRH:LTP07,CDH:PRB13} in Tables \ref{tab:N36}--\ref{tab:N63}. 
The column quoting the total number of loop configurations is not very 
relevant for our present purposes, but it demonstrates the large number 
of allowed loop configurations, in particular in the single-loop sector 
($n_\ell=1$).

Firstly, for $N=27$, $N=45$, and $54$ there are $N/9$ more 6-loop 
configurations than hard hexagons in the sector $n_\ell =1$. Indeed, as 
already stated in Sec.~\ref{sec:latt}, one can see in 
Fig.~\ref{magnon_crystal-lattice} that there are loops of length 6 
winding around the boundary. This implies also the existence of a fourth 
``magnon crystal'' in the sector with $n_\ell = N/9$ consisting of these 
loops winding around the boundary, thus explaining why the degeneracy of 
the Heisenberg model in the sector with $S^z=35/2$ and $21$ respectively 
is found to be $4$ on the $N=45$ and $54$ lattices, and not just $3$ as 
expected from hard hexagons, see Tables \ref{tab:N45} and \ref{tab:N54}.

Secondly, for most sectors with $n_\ell \ge 3$ the number of linearly 
independent loop configurations is smaller than the degeneracy of the 
corresponding sector of the Heisenberg model. This includes the sector 
$S^z=14$ on the $N=36$ lattice where the degeneracy is $8$ while there 
are only $3$ configurations consisting of $4$ loops (see Table 
\ref{tab:N36}). Consequently, some ground states of the Heisenberg model 
are evidently not captured by the loop picture; the biggest difference 
that we have observed is $648$ non-loop ground states of the Heisenberg 
model in the sector $S^z=31$, $n_{\ell} = 5$ on the $N=72$ lattice, see 
Table~\ref{tab:N72}. However, there are significantly more linearly 
independent loop configurations than simple hard-hexagon configurations 
such that the present picture amounts to an improved description of the 
ground-state manifold of the Heisenberg model.

There is a third intriguing observation pointing towards a significant 
difference between the $N=63$ and $N=72$ lattices. For $N=63$ and in the 
sector $n_{\ell} = N/3-1 =6$, there are exactly as many loop as 
hard-hexagon configurations, leaving no room for linear relations, see 
Table \ref{tab:N63}. On the other hand, Table \ref{tab:N72} shows that in 
the corresponding case $n_{\ell} = N/3-1 =7$ for $N=72$, there are more 
than 3 times as many linearly independent loop as hard-hexagon 
configurations.

Figure 4(a) of the main manuscript characterizes the total ground-state 
degeneracy in terms of the associated entropy per site. We should mention 
that the $N=72$ `ED' data point in this figure is only a lower bound 
since we were so far not able to compute the ground-state degeneracy of 
the Heisenberg model in the sectors with $S^z \le 30$, as specified by 
the missing entries in Table~\ref{tab:N72}, such that in these cases we 
have used the number of linearly independent loop configurations as a 
lower bound.

\subsection{Specific heat}

\begin{figure*}[tb!]
\centering
\includegraphics[width=7 true cm]{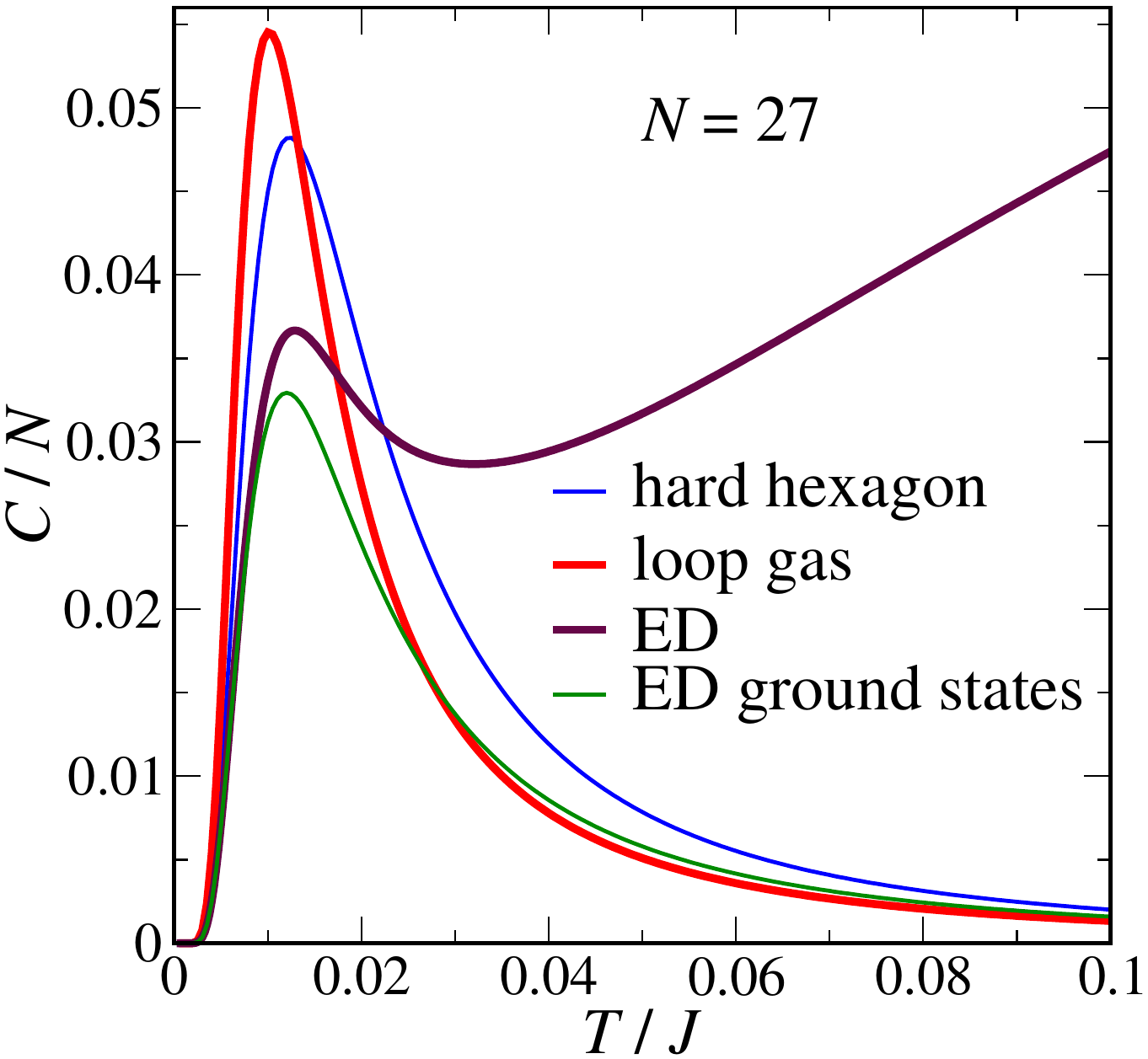}\qquad%
\includegraphics[width=7 true cm]{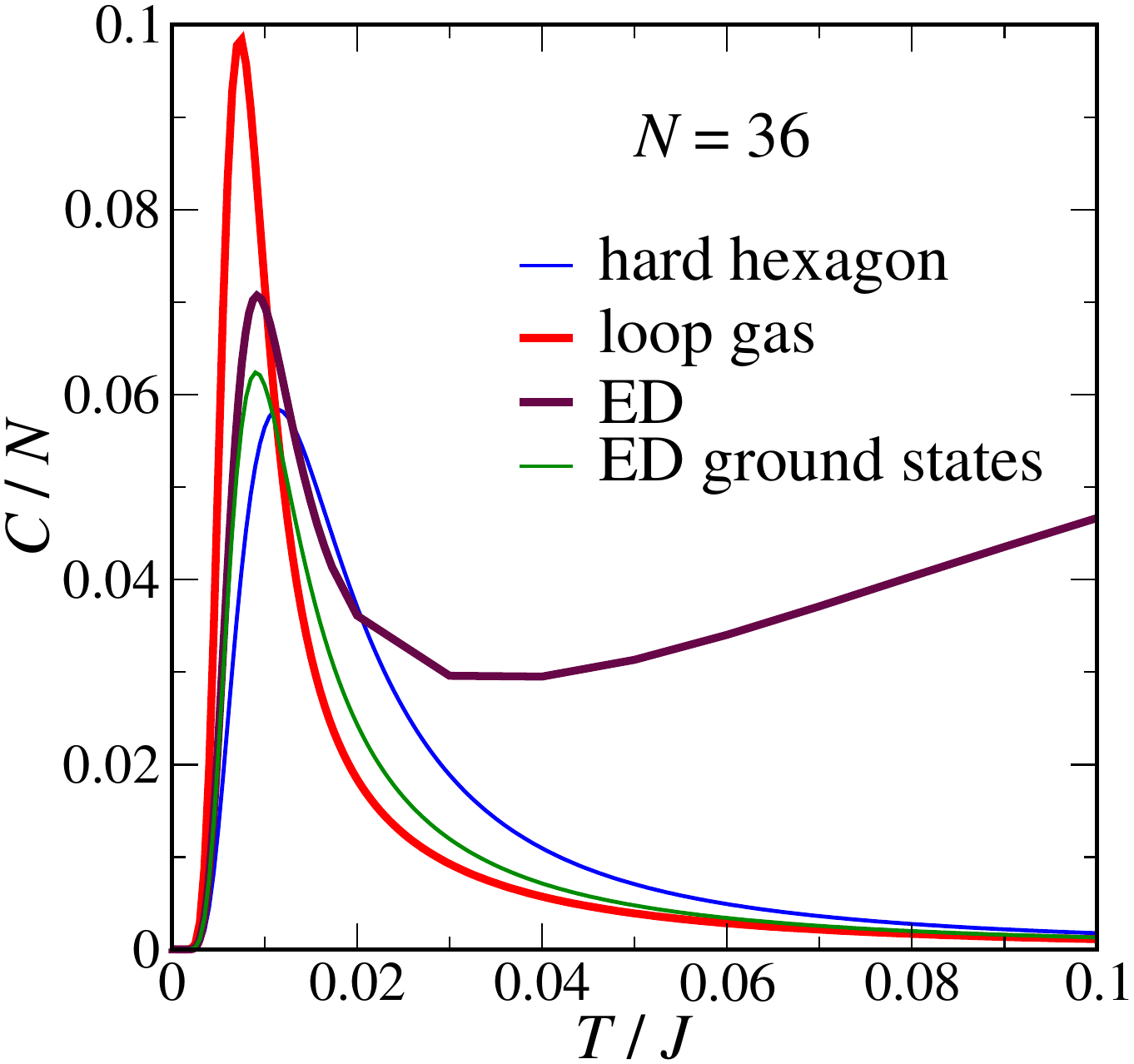} \\
\includegraphics[width=7 true cm]{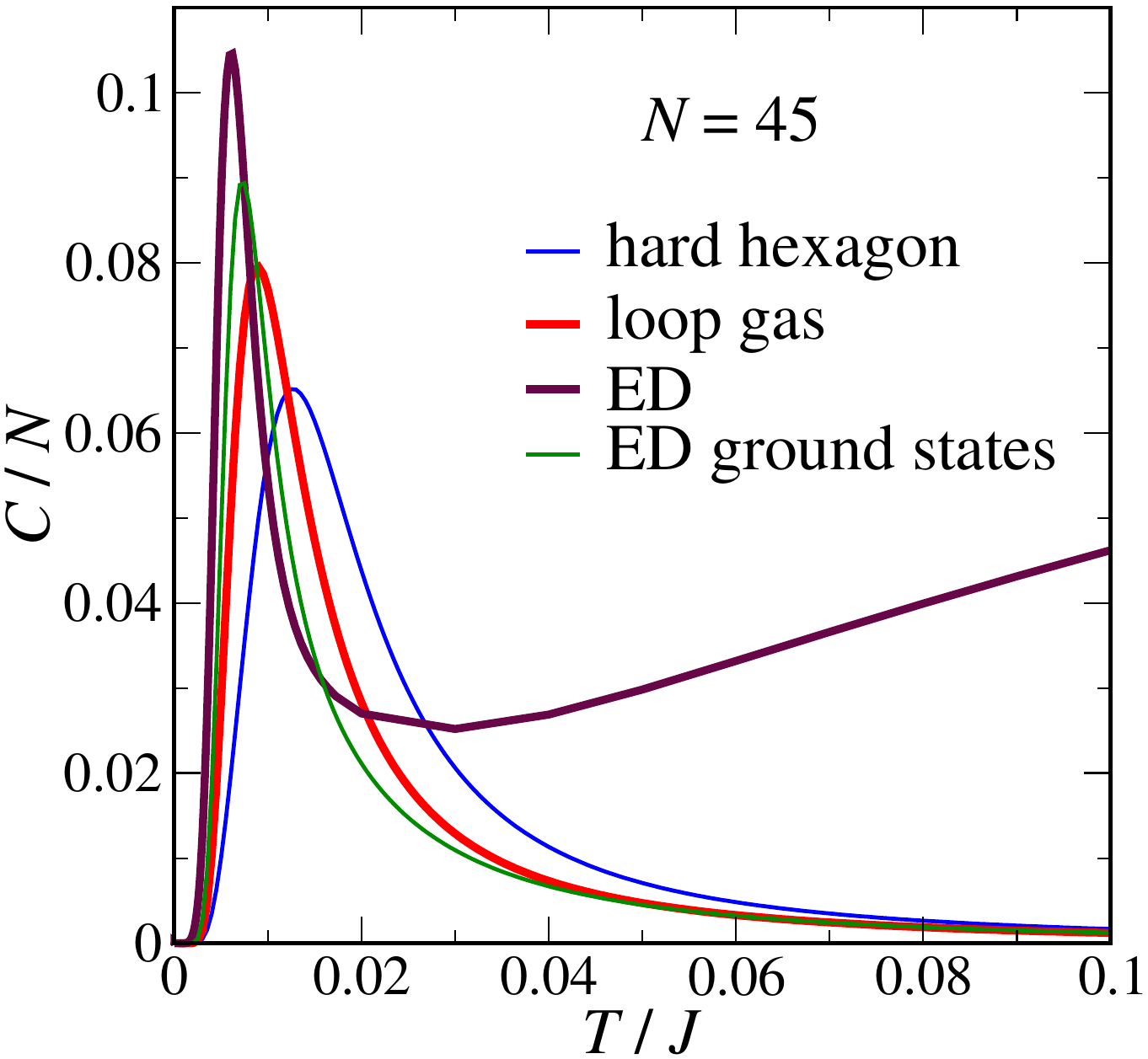}\qquad%
\includegraphics[width=7 true cm]{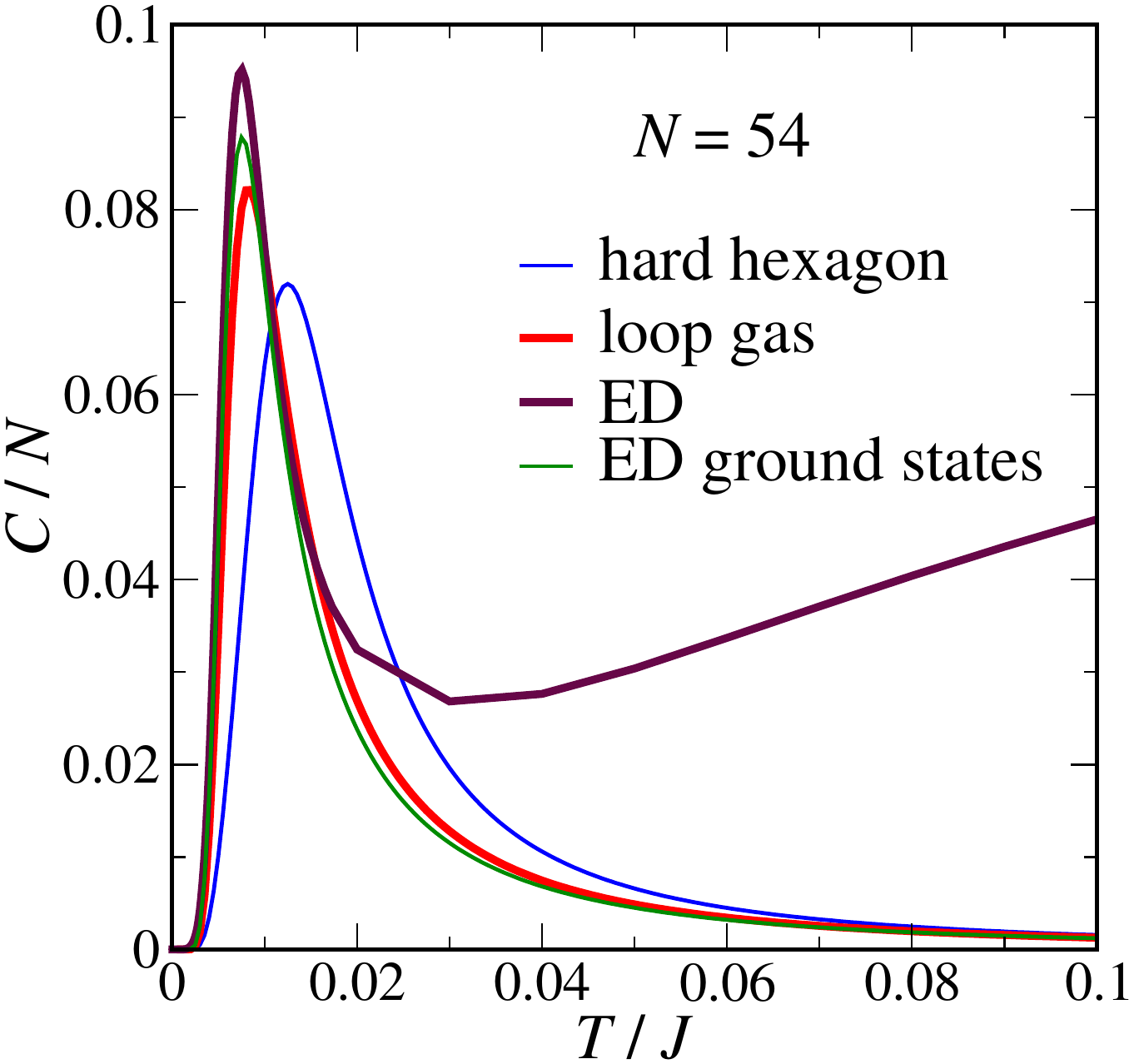} \\
\includegraphics[width=7 true cm]{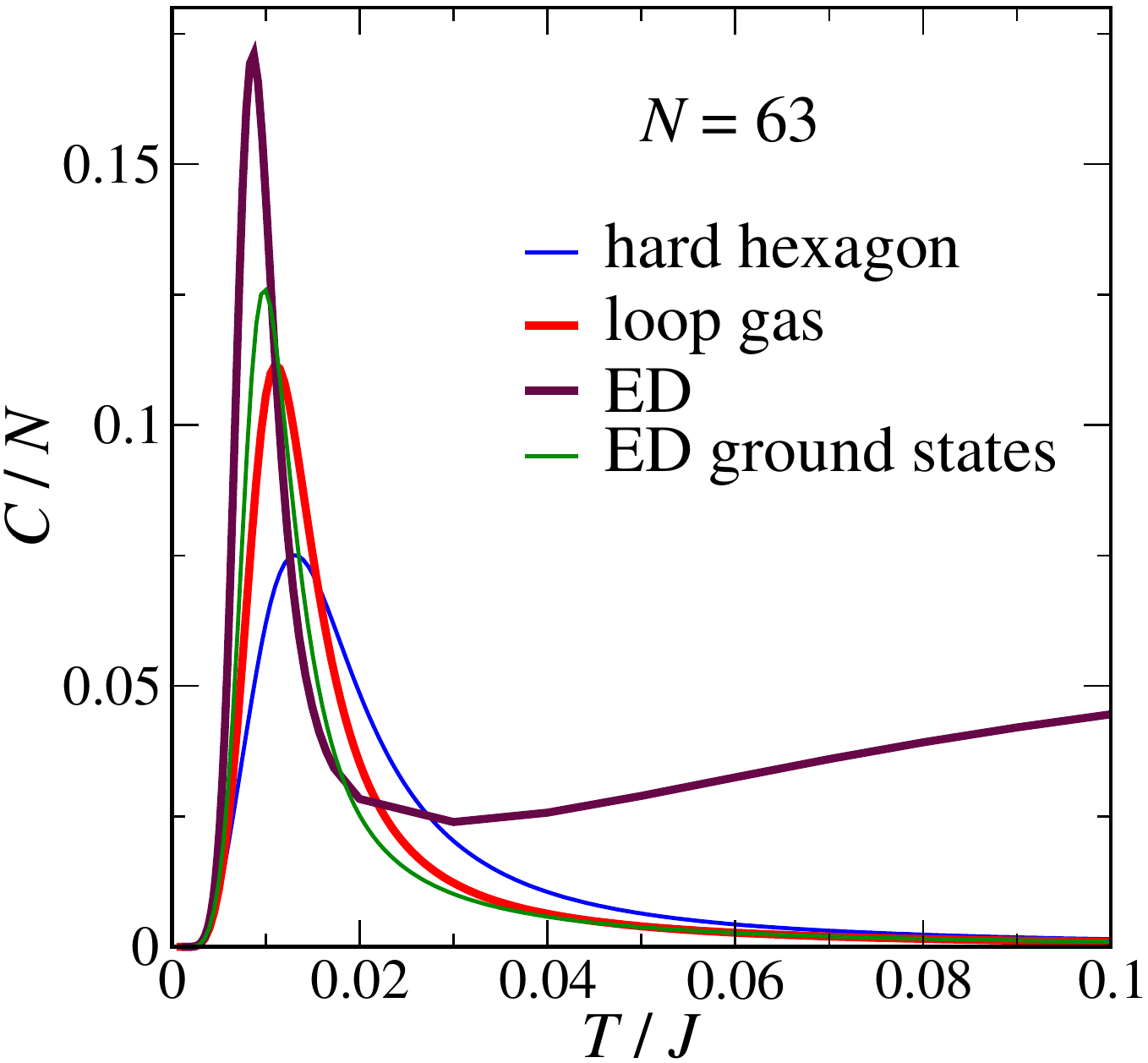}\qquad%
\includegraphics[width=7 true cm]{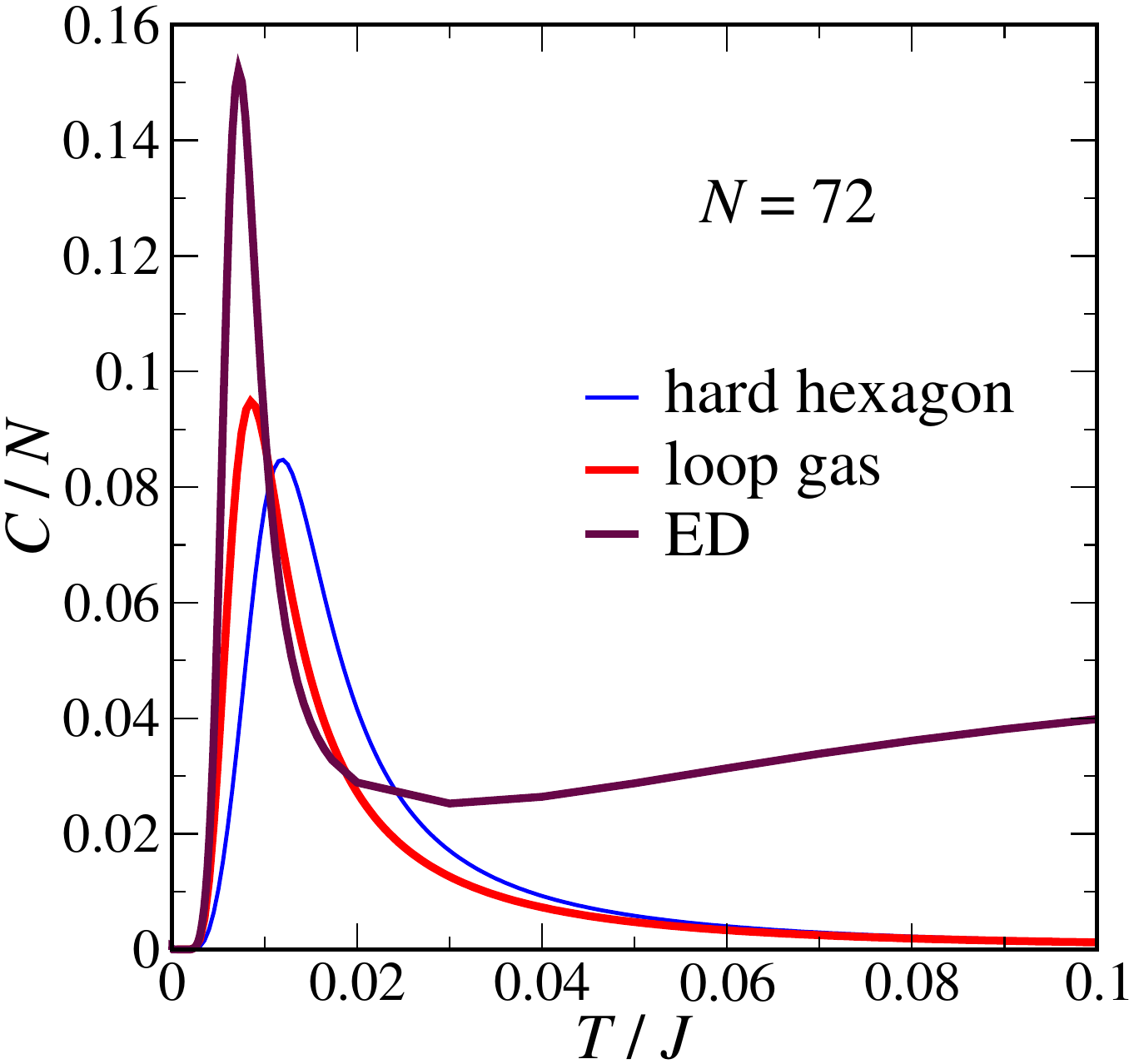}
\caption{Specific heat per site $C/N$ for the $N=27$, $36$, $45$, $54$, $63$, and $72$ lattices at $B=0.99\,B_{\rm sat}$.
The green lines in the $N=27,\ldots,63$ panels correspond to just taking the ground states
of the Heisenberg model into account whose numbers are given in Tables~\ref{tab:N27}--\ref{tab:N63}.
\label{fig:CN}
}
\end{figure*}

\begin{figure*}[tb!]
\centering
\includegraphics[width=7 true cm]{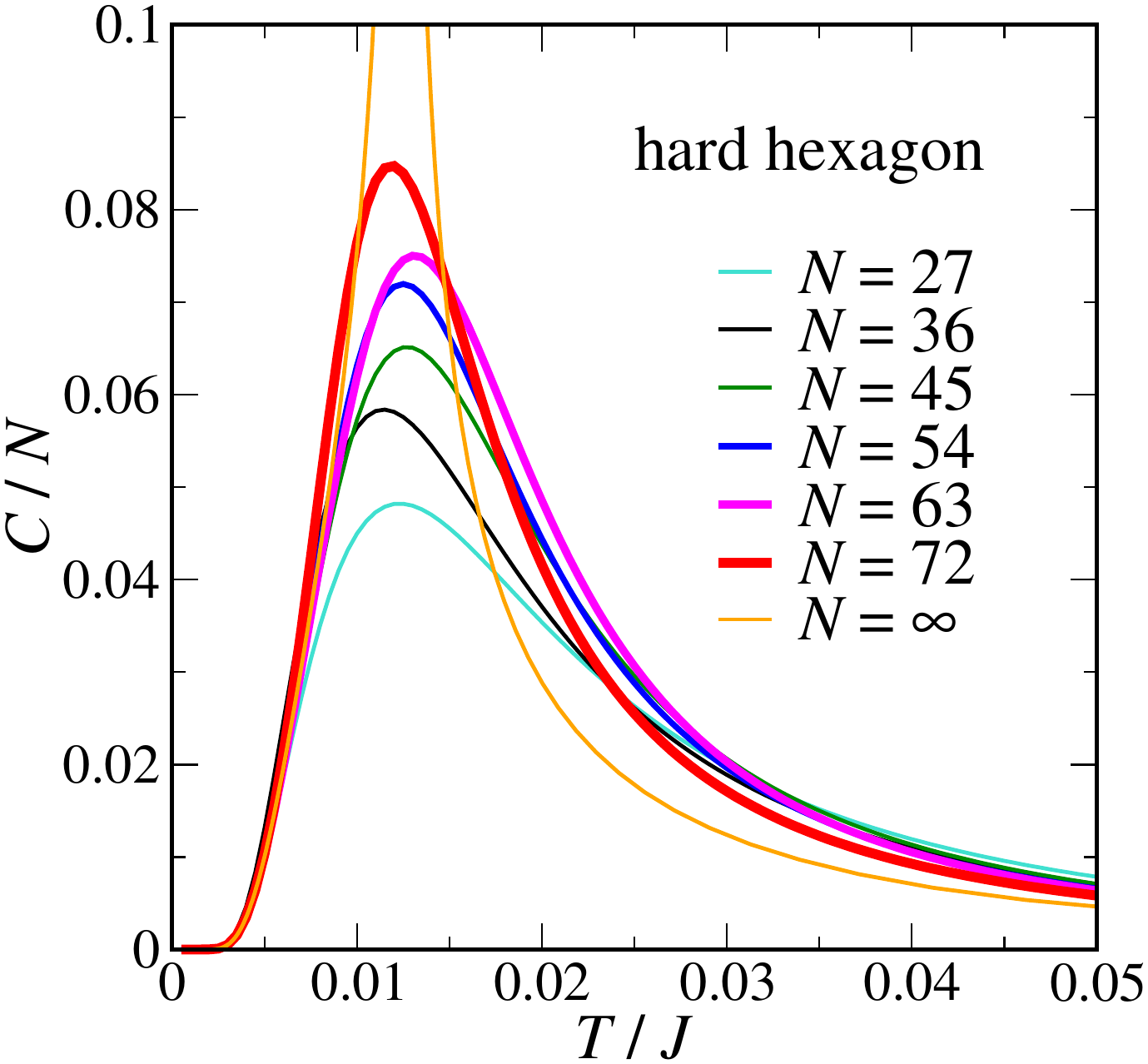}\qquad
\includegraphics[width=7 true cm]{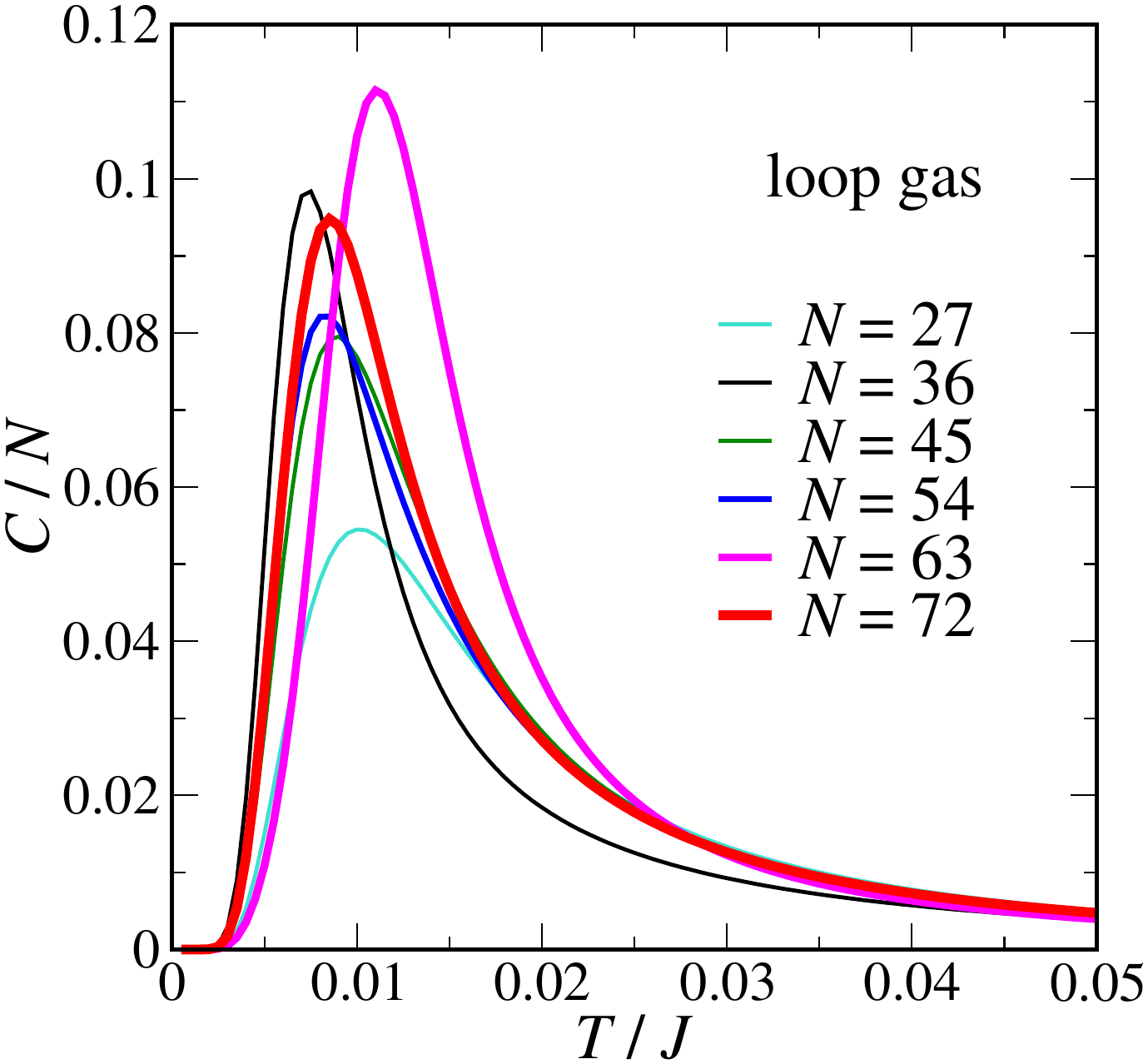}
\\
\includegraphics[width=7 true cm]{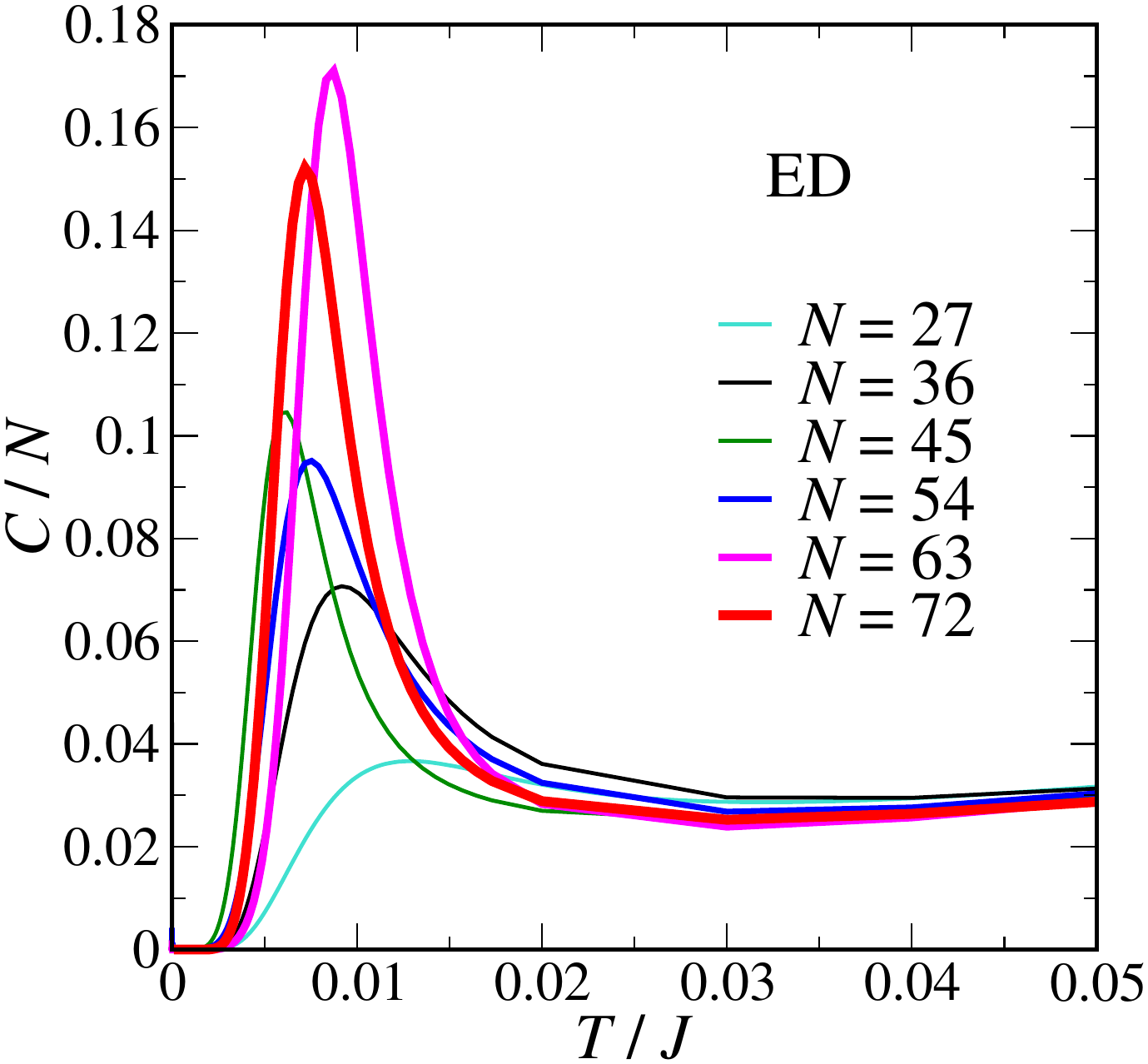}
\caption{Same data as in Fig.~\ref{fig:CN}, but grouped
  according to method rather than size $N$. The hard-hexagon
  panel includes also the result for the thermodynamic 
limit \cite{ZhT:PRB04,Bax:JPA80,ZhT:PTPS05}.
\label{fig:Cmethod}
}
\end{figure*}

{}From the multiplicities in Tables~\ref{tab:N27}--\ref{tab:N72} it is 
straightforward to compute thermodynamic properties such as the specific 
heat. We note that both for hard hexagons and the loop gas, the results 
depend only on $(B_{\rm sat}-B)/T$ which plays the role of a chemical 
potential \cite{DeR:EPJB06,ZhT:PRB04,ZhT:PTPS05}. Nevertheless, for 
comparison with exact diagonalization (ED), we take a value of the 
magnetic field close to the saturation field, $B= 0.99 \, B_{\rm sat}$. 
The results for the specific heat per site $C/N$ are shown in 
Fig.~\ref{fig:CN} (here we set $k_{\rm B}=1$). This figure also includes 
the exact diagonalization results for the Heisenberg model from the main 
text. Although the behavior is somewhat irregular as a function of system 
size $N$, the loop-gas description is evidently in better agreement with 
the full Heisenberg model than the simpler hard-hexagon model. This 
concerns in particular the position of the low-temperature maximum where 
the loop gas and the Heisenberg model yield almost identical positions 
for $N=54$ and $72$. In order to emphasize this point even more, we show 
in the $N=27,\ldots,63$ panels in green the specific heat that is 
obtained when we take into account just the ground-state degeneracy of 
the Heisenberg model according to Tables~\ref{tab:N27}--\ref{tab:N63}. 
Firstly, we see that the low-temperature peak of the specific heat is 
indeed dominated by the ground-state manifold, and that the loop gas 
yields in turn a rather accurate account of these. Nevertheless, one 
observes also that the full ED result for the low-temperature peak of the 
specific heat is clearly higher even than just the ground-state 
contribution. This can be understood from the gap $\Delta$ to the first 
excited state that is also included Tables~\ref{tab:N27}--\ref{tab:N72}. 
Indeed, this gap is so small, at least in some sectors, that even at a 
temperature as low as $T \approx J/100$ the contribution from thermally 
excited states to the specific heat is quantitatively relevant.

Generally, the maximum value of the specific heat remains smaller in the 
loop gas than in the full Heisenberg model on the \kagome\ lattice (see 
also Fig.~4(c) of the main text). This is consistent with the number of 
linearly independent loop-gas states being lower than the number of 
independent ground states, see Tables~\ref{tab:N27}--\ref{tab:N72} and 
Fig.~4(a) of the main text. The exceptions to this general rule are the 
cases $N=27$ and $36$. The latter can be traced back to a difference of 
the number of ground states that for the case $N=36$ are in the sector 
$S^z=14$, $n_\ell=4$: for the full Heisenberg model, this degeneracy is 8 
while there are only 3 linearly independent loop-gas and hard-hexagon 
configurations, see Table~\ref{tab:N36}. A similar difference appears for 
$N=27$, namely 13 versus 6, respectively 3 in the sector $S^z=21/2$, 
$n_\ell=3$, see Table~\ref{tab:N27}. Consequently, the leading term of 
the low-temperature expansion of the partition function $Z$ is very 
different, thus giving rise to the significant differences in the 
low-temperature specific heat $C/N$ that one observes in the panels for 
$N=27$ and $36$ of Fig.~\ref{fig:CN}.

Figure \ref{fig:Cmethod} regroups this data for the specific heat 
according to method in order to expose the finite-size behavior more 
clearly (another presentation of the ED results is given in Fig.~3 of the 
main text). One sees in Fig.~\ref{fig:Cmethod} that the hard-hexagon 
description yields a maximum that increases monotonically with increasing 
$N$ (see also Fig.~4(c) of the main text). Here, the thermodynamic limit 
is known \cite{ZhT:PRB04,Bax:JPA80,ZhT:PTPS05} such that we can include 
this result in the hard-hexagon panel. By contrast, both the loop-gas 
description and ED yield significantly less regular behavior. However, 
the changes as a function of $N$ are very similar for the latter two 
methods, in particular when one goes from $N=63$ to $72$, the notable 
exceptions being again the cases $N=27$, $36$.

Overall, we conclude that the loop-gas description yields not only a 
substantial improvement over hard hexagons, but also that the $N=63$ and 
$72$ lattices are more representative of the generic behavior than the 
smaller lattices, thus underlining again also the importance of being 
able to access such big system sizes by FTL.


%